\documentclass[12pt]{iopart}

\usepackage[T1]{fontenc}

\usepackage[latin1]{inputenc}

\usepackage{graphicx}

\usepackage{multirow}

\usepackage[english]{babel}

\usepackage{amssymb}

\usepackage{amsfonts}

\usepackage{subfig}

\usepackage{epsfig}

\newcommand{\be}{\begin{equation}}

\newcommand{\ee}{\end{equation}}

\newcommand{\bea}{\begin{eqnarray}}

\newcommand{\eea}{\end{eqnarray}}

\newcommand{\beaa}{\begin{eqnarray*}}

\newcommand{\eeaa}{\end{eqnarray*}}

\begin{document}

\title[On the stability in $f(R,G)$ gravity]{On the stability of the cosmological solutions in $f(R,G)$ gravity}
\author{\'Alvaro de la Cruz-Dombriz$^{1,2}$ and Diego S\'aez-G\'omez$^{3}$}
\address{$^{1}$ Astrophysics, Cosmology and Gravity Centre (ACGC), University of Cape Town, 7701 Rondebosch, Cape Town, South Africa}
\address{$^{2}$ Department of Mathematics and Applied Mathematics, University of Cape Town, 7701 Rondebosch, Cape Town, South Africa}
\address{$^{3}$ Fisika Teorikoaren eta Zientziaren Historia Saila, Zientzia eta Teknologia Fakultatea, Euskal Herriko Unibertsitatea, 644 Posta Kutxatila, 48080 Bilbao, Spain, EU}

%
%
%
%
\begin{abstract}
Modified gravity is one of the most promising candidates for explaining the current accelerating expansion of the Universe, and even its unification with the inflationary epoch. Nevertheless, the wide range of models capable to explain the phenomena of dark energy, imposes that current research focuses on a more precise study of the possible effects of modified gravity may have on both cosmological and local levels. In this paper, we focus on the analysis of a type of modified gravity, the so-called $f(R,G)$ gravity and we perform a deep analysis on the stability of important cosmological solutions. This not only can help to constrain the form of the gravitational action, but also facilitate a better understanding of the behavior of the perturbations in this class of higher order theories of gravity, which will lead to a more precise analysis of the full spectrum of cosmological perturbations in future.  
\end{abstract}

\maketitle

\section{Introduction}

In the recent years, {\it modified gravity theories} have become one of the most popular candidates for explaining the current accelerated expansion of the Universe. As it is very well known,  general relativity (GR) in its 
standard form can not explain such behavior of the Universe expansion without either extra terms in the gravitational Lagrangian (for  reviews on modified theories of gravity, see Refs.~\cite{review,book}) or exotic fluid components (see Refs.~\cite{Ref1}-\cite{Ref3}).
Modified gravity theories have been widely studied, and it has been shown that they are not only capable to mimic the dark energy epoch, but also the inflationary era \cite{unification}. 
Therefore, by the only use of large scale observations (Ia type supernova, BAO, or the cosmic microwave background) which depend uniquely on the evolution history of the Universe,  the nature and the origin of DE cannot be determined due to the fact that identical evolutions for the cosmological background can be explained by a diverse number of theories. This is the so called
{\it degeneracy problem}. It is thus required, in order to confirm or discard the validity of these theories, to obtaining solutions that can also describe correctly, e.g., 
the growth factor of scalar perturbations (see Refs.~\cite{Perturbations}), the stability of cosmological solutions against small perturbations 
and the existence of GR-predicted astrophysical objects such as black holes \cite{BH}.  \\

In this sense the simplest, and in fact the most studied, modification of GR is the one where the Hilbert-Einstein action is generalized to a general function of the Ricci scalar $R$, the so-called $f(R)$ gravities  \cite{Ref5}-\cite{Ref6}. These theories are able to mimic the behavior of the cosmological constant (see for instance Ref.~\cite{gr-qc/0607118}), but also can reproduce the entire cosmological history (see Ref.~\cite{unification}).  In addition, they seem to behave quite well at local scales, where the GR limit must be recovered \cite{Viable}, and the existence of GR-predicted astrophysical objects such as black holes can be achieved 
\cite{BH_Dombriz}. Nonetheless, these candidates have their own shortcomings \cite{Ref6} and have to pass rigorous theoretical and observational scrutiny before they can be accepted as viable theories \cite{Viable}. 
Another possible modification of the standard gravitational Lagrangian includes a wider number of curvature invariants ($R$, $R_{\mu\nu}R^{\mu\nu}, R_{\mu\lambda\nu\sigma}R^{\mu\lambda\nu\sigma}$ among others). Within these modifications, the so-called Gauss-Bonnet gravity can be included. In these theories the gravitational Lagrangian consists of a function $f(R,G)$ where $G$ holds for the usual Gauss-Bonnet invariant. 
This modified gravity has been also widely studied and it is known that can also reproduce any kind of cosmological solution (see Refs.~\cite{Cog1}-\cite{N2}), where special attention has been already paid to models able to mimic the $\Lambda$CDM model, as well as other important cosmological solutions (see Refs.~\cite{arXiv:1001.3636, arXiv:1009.0902}). Finally, the cosmological perturbations have been explored within different standard scenarios for this class of theories \cite{Felice_malo}. \\

In this investigation, we are interested in studying the stability behavior of several kind of cosmological solutions in the framework of Gauss-Bonnet gravities when subjected to homogeneous perturbations.
%
Our analysis will therefore exclude anisotropic, i.e. cosmological scalar, perturbations.  Homogeneous and isotropic perturbations 
have been historically considered as the mean to determine the stability of different modified gravity theories (see for instance Refs. \cite{Ref1,Barrow, arXiv:1011.2090}).
The usual approach in these references consisted in perturbing both the Hubble parameter and the matter density in order
to check the background stability in the time
evolution. This way, one can determine the stability of the modified
Einstein equations when treated as differential equations. With
respect of the cosmological perturbations, the full anisotropic analysis 
of cosmological perturbations in modified gravity theories is out of the scope of this 
investigation whereas some significant advances have been made in the last years, in particular for 
$f(R)$ theories \cite{Perturbations}.\\



With regard to Gauss-Bonnet gravity theories,
a particular subclass has been studied in Refs.~\cite{arXiv:0810.5712,arXiv:0911.1811}, where the Hilbert-Einstein action plus a function $f(G)$ is considered. However,  the extension to a more general form for the $f(R,G)$ gravitational theory is a mandatory task that may help to understanding the viability and features of more general Lagrangians. 
These theories may present ghost degrees of freedom in an empty anisotropic universe, i.e. the Kasner-type background \cite{Felice_peor}. Nonetheless, these degrees of freedom are absent on Friedmann-Lema\^itre-Robertson-Walker (FLRW) backgrounds (see also \cite{Felice_peor}). This is precisely the type of cosmological background that will be considered in our investigation. 
Hence, in the present paper we analyze some important cosmological solutions in $f(R,G)$ gravity, both in a vacuum scenario and with the presence of standard perfect fluids. The stability of cosmological solutions has been studied for $f(R)$ gravity in Ref.~\cite{Barrow}, as well as for other curvature invariants in Ref.~\cite{Ref1}. For some particular cases of $f(R,G)$, the perturbations on the solutions have been analyzed in Ref.~\cite{Saltas:2010tt}, whereas for Ho\v{r}ava-Lifshitz gravity, the analysis has been performed in Ref.~\cite{arXiv:1011.2090}. Here we extend the analysis to more general actions, where we shall find the stability conditions for different cosmological evolutions in FLRW universes such as inflationary epoch and late-time accelerated era as described by the $\Lambda$CDM model. 
Therefore, these analyses can help understanding the viability of cosmological evolution provided by this kind of modified gravity, and constraining the viable 
candidates for the underlying gravitational action. \\


The paper is organized as follows:
 in Section II we present the general features of the $f(R,G)$ gravity theories 
 by writing the corresponding modified Einstein equations. In Section III 
 we introduce the evolution equations of perturbations appearing 
 in these scenarios once that a FLRW background is assumed. Sections IV  and V are then devoted to the study of stability around the de Sitter and 
 power-law solutions respectively. In the last case, we pay special attention to configurations including perfect fluids such as radiation and dust.  
 Section VI is finally devoted to study the stability of the $f(R,G)$ model able to mimic the $\Lambda$CDM cosmological evolution without any cosmological constant. 
We conclude the paper by giving our conclusions in Section VII. An appendix is included at the end of the communication to provide
explicitly the coefficients of perturbations equations introduced in Section III.

\section{$F(R,G)$ gravity}

Let us start by writing the most general action for modified Gauss-Bonnet gravity, which is given by, 
\begin{equation}
S=\int d^{4}x\sqrt{-g}\left[\frac{1}{2\kappa^2}f(R,G)+\mathcal{L}_{m}\right].
\label{1.1}
\end{equation}
where $\kappa^2=8\pi G_N$, $G_N$ is the Newton constant and $\mathcal{L}_m$ represents the matter Lagrangian.  
The symbol $G$ holds for the Gauss-Bonnet invariant, which is expressed as,
\begin{equation}
 G\,\equiv\,R^2-4R_{\mu\nu}R^{\mu\nu}+R_{\mu\nu\lambda\sigma}R^{\mu\nu\lambda\sigma}
\label{GB}
\end{equation} 
Then, by varying expression (\ref{1.1}) with respect to the metric tensor $g_{\mu\nu}$, the modified Einstein field equations are obtained \cite{Cog1},
$$
0\,=\,\kappa^{2} T^{\mu\nu}+\frac{1}{2}g^{\mu\nu}f(G)-2f_{G}RR^{\mu\nu}+4f_{G}R^{\mu}_{\rho}R^{\nu\rho}
$$
$$
-2f_{G}R^{\mu\rho\sigma\tau}R^{\nu}_{\rho\sigma\tau}-4f_{G}R^{\mu\rho\sigma\nu}R_{\rho\sigma}+2(\nabla^{\mu}\nabla^{\nu}f_{G})R
-2g^{\mu\nu}(\nabla^{2}f_{G})R-4(\nabla_{\rho}\nabla^{\mu}f_{G})R^{\nu\rho}
$$
$$
-4(\nabla_{\rho}\nabla^{\nu}f_{G})R^{\mu\rho}+4(\nabla^{2}f_{G})R^{\mu\nu}+4g^{\mu\nu}(\nabla_{\rho}\nabla_{\sigma}f_{G})R^{\rho\sigma}
$$
\begin {equation}
-4(\nabla_{\rho}\nabla_{\sigma}f_{G})R^{\mu\rho\nu\sigma}-f_{G}R^{\mu\nu}+\nabla^{\mu}\nabla^{\nu}f_{R}-g^{\mu\nu}\nabla^{2}f_{R}.
\label{1.2}
\end{equation}
where $\nabla$ holds for the usual covariant derivative and subindices $G$ and $R$ in $f$ hold for derivatives of the gravitational Lagrangian $f(R,G)$ with respect to those arguments.
In this investigation, we are interested in studying different cosmological solutions described by a flat FLRW Universe, such that we assume along the paper the metric, 
\be
ds^2\,=\,-dt^2+a^2(t)\sum_{i=1}^{i=3}\left(d\ x^{i}\right)^{2}
\label{1.3}
\ee
Then, the Hubble parameter $H$ takes its usual definition $H\equiv \dot{a}/a$, and $G$ and $R$ become
\begin {equation}
G=24(\dot{H}H^{2}+H^{4}),\quad R=6(\dot{H}+2H^{2}),
\label{1.4}
\end{equation}
where the dot holds for the derivative with respect to cosmic time $t$. By substituting expressions (\ref{1.4}) and the metric (\ref{1.3}) into the field equations (\ref{1.2}), the FLRW equation for indices $\mu=\nu=0$ yields
\begin {equation}
0=\frac{1}{2}(Gf_{G}-f-24H^{3}\dot{f}_{G})+3(\dot{H}+H^{2})f_{R}-3H\dot{f}_{R}+\kappa^2\rho_{m}.
\label{1.4_back}
\end{equation}
Let us assume from now on, that the matter fluid will be given under the form of a 
perfect fluid with a constant equation of state $p_m=\omega\rho_{m}$, with the matter energy density 
$\rho_{m}$ satisfying the standard continuity equation
\begin {equation}
\dot{\rho}_{m}+3H(1+w)\rho_{m}=0.
\label{1.5}
\end{equation}

\section{Perturbations of flat FLRW solutions in $F(R,G)$ gravity}

Let us now study the homogeneous and isotropic perturbations around a particular cosmological solution for the $f(R,G)$ theory described by the action (\ref{1.1}). Here we establish the perturbed equations for a general case, but specific cases will be studied in the upcoming sections, specially de Sitter and power law solutions, as well as the behavior of $\Lambda$CDM solution 
in the context of these theories of gravity. 

For this purpose, let us assume a general solution for the cosmological background of FLRW metric, which is given by a Hubble parameter 
$H=H_0(t)$
that 
satisfies the background equation (\ref{1.4_back}) for a particular $f(R,G)$ model. The evolution of the matter energy density can be expressed in terms of this particular solution by solving the continuity equation (\ref{1.5}) yielding
\be
\rho_{m0}(t)=\rho_0\,\e^{-3(1+w_m)\int H_0(t){\rm d}t}\ .
\label{2.2}
\ee
%
Since we are interested in studying the perturbations around the solutions $H=H_0(t)$ and density given by (\ref{2.2}), let us consider small deviations from the Hubble parameter and the energy density evolution. Hence,
\be
H(t)=H_0(t)\left(1+\delta(t)\right)\ , \quad \rho_{m}(t)=\rho_{m0}(t)\left(1+\delta_m(t)\right)\ .
\label{2.3}
\ee
where $\delta(t)$ and $\delta_{m}(t)$ hold for the isotropic deviation of the background Hubble parameter and the matter overdensity respectively.
In order to study the behavior of these perturbations in the linear regime, we expand the function $f(R,G)$ in powers of $R_0$ and $G_0$ evaluated at the solution $H=H_0(t)$,
\be
f(R,G)\,=\,f^0+f_R^0(R-R_0)+f_G^0(G-G_0)+\mathcal{O}^2
%
\label{2.4}
\ee
where the superscript $0$ refers to the values of $f(R,G)$ and its derivatives evaluated at $R=R_0$ and $G=G_0$. 
The $\mathcal{O}^2$  term includes all the terms
proportional to square or higher powers of $R$ and $G$ that will be included in the equation, although only the linear terms of the induced perturbations are considered. 
Hence, by introducing expression (\ref{2.3}) in the FLRW background equation (\ref{1.4_back}) and using the expansion (\ref{2.4}), the equation for the perturbation $\delta(t)$ becomes in the linear approximation,
\be
c_2\ddot{\delta}(t)+c_1\dot{\delta}(t)+c_0\delta(t)=c_m\delta_m(t)\ ,
\label{2.5}
\ee
where coefficients $c_{0,1,2}$ and $c_m$ are explicitly written in the Appendix at the end of the communication. These coefficients depend explicitly on the $f(R,G)$ and its derivatives evaluated in the background solution. In addition, there is a second perturbed equation obtained from the matter continuity equation (\ref{1.5}) once it is perturbed with expressions (\ref{2.3}). Thus,
\be
\dot{\delta}_m(t)+3H_0(t)\delta(t)=0\ .
\label{2.6}
\ee
Hence, for a particular FLRW cosmological solution, its stability can be studied in the context of $f(R,G)$ gravity by analyzing and solving the equations (\ref{2.5}) and (\ref{2.6}). Due to the linear character of (\ref{2.5}), the solution for $\delta(t)$ can be in general split in two branches: the first one corresponding to the solution of the homogeneous equation in (\ref{2.5}), which reflects the perturbations induced by the chosen particular gravitational Lagrangian. The second branch would correspond to the particular solution of that equation, which is merely affected by the growth of matter perturbations $\delta_{m}(t)$. Hence, the general solution can be written as,
\be
\delta(t)=\delta_{homogeneous}(t)+\delta_{particular}(t)\ .
\label{2.7}
\ee
 In the upcoming sections, we shall consider several cosmological solutions, and their stability will be then studied. Nevertheless, let us firstly do some general considerations. By recovering the Hilbert-Einstein $f(R,G)=R$, the stability equation (\ref{2.5}) yields,
\be
-6 H_0^2\delta(t)=c_m\delta_m(t)\ .
\label{2.8}
\ee
which can be understood as an algebraic relation between the geometrical and the matter perturbations.
Hence, in GR the full perturbation around a cosmological solution is fully determined by the matter perturbations (or vice versa). In fact, by taking explicit expression for $c_m$ in the Appendix and equation (\ref{2.6}) it is straightforward to prove that
\be
\delta(t)\,=\,-\frac{1}{2}\delta_{m}(t)\,\propto\,a(t)^{3/2}
\label{GR}
\ee

Nevertheless, the algebraic relation (\ref{GR}) between $\delta(t)$ and $\delta_m(t)$ in GR is in general absent for higher order theories of gravity. For these theories, the evolution of the perturbations is in general determined by a coupled system of ordinary differential equations,  (\ref{2.5}) and (\ref{2.6}), where the underlying gravitational Lagrangian plays an essential role on the form of the $c_{0,1,2}$ coefficients as can be seen in the Appendix.
Let us illustrate the previous comments by considering theories whose Lagrangians can be rewritten as follows
\be
f(R,G)=f_1(G)+f_2(R)\ .
\label{2.9}
\ee
Then, the perturbation equation (\ref{2.5}) becomes much simpler and some information can consequently be extracted. For instance, after assuming that GR should be recovered in some limit, which basically means that the higher derivatives of the function (\ref{2.9}) should become negligible, the coefficient $c_1$ in (\ref{2.5}) becomes null, and by diving the equation by $c_2$, it yields,
\be
\ddot{\delta}(t)+\frac{f^0_R}{3(16f^{0}_{GG}H_{0}(t)^4+f^{0}_{RR})}\delta (t)\,=\,  \frac{c_m}{18(16f^{0}_{GG}H_{0}(t)^4+f^{0}_{RR})} \,\delta_m(t)\ .
\label{2.10} 
\ee
where $c_m'=\frac{c_m}{18(16f^{0}_{GG}H_{0}(t)^4+f^{0}_{RR})}$. In order to ensure the stability of a particular solution in vacuum in the GR limit and provided that $f_{R}^{0}>0$, the denominator in (\ref{2.10}) has to satisfy,
\be
16f^{0}_{GG}H_{0}(t)^4+f^{0}_{RR}>0\ .
\label{2.11}
\ee
In fact, this constraint was also proved in \cite{GBstability} to guarantee the generalized second law of Thermodynamics for these theories in de-Sitter scenarios.
Note that for $f(R,G)=R+f (G)$, the Lagrangian is restricted to be $f^{0}_{GG}>0$ to ensure the stability of any solution in the GR limit  \cite{arXiv:0810.5712}. Nonetheless, this constraint on the second derivative with respect to $G$ may produce, according to \cite{arXiv:0911.1811}, instabilities in the matter cosmological perturbations for this kind of models. However, a wider range of functions $f_1(R)$ in (\ref{2.9}) may circumvent the existence of such instabilities.

\section{Stability of De Sitter solutions}

Let us start our study of the stability of different cosmological solutions by studying some of the simplest 
cosmological solutions, namely the de-Sitter (dS) solutions,
\be
H_{0}(t)=H_0\quad \rightarrow \quad a(t)=a_0\e^{H_0t}\ ,
\label{dS1}
\ee
where $H_0$ is constant. According to (\ref{1.4}), the Ricci and Gauss-Bonnet terms are given in this case by $R_0=12H_0^2$ and $G_0=24H_0^2$. Then, inserting the expression for the Hubble parameter (\ref{dS1}) in equation (\ref{1.4_back}) once vacuum is considered, we get
\be
\frac{1}{2}\left(G_0f_G(R_0,G_0)-f(G_0,R_0)\right)+3H_0^2f_R(R_0,G_0)=0
\label{dS2}
\ee 
Therefore, any $f(R,G)$ function can in principle admit vacuum de Sitter solutions provided that the previous algebraic equation has positive roots for $H_0$. Hence, for some particular $f(R,G)$ models, the current accelerating epoch of the Universe expansion -- as well as the inflationary epoch -- can be explained.   Following a different approach, one can consider equation (\ref{dS2}) as an ordinary differential equation for the $f(R_0,G_0)$ function so that the corresponding solution would admit any $H_0^2$ value. Thus, by solving the equation (\ref{dS2}) in terms of $f(R,G)$, the following action is obtained,

\be
f(R,\,G)\,=\,\alpha G \left[R - 6 H_0^{2}\log(G)\right]
\label{eqn_nueva}
\ee
where $\alpha$ is an arbitrary integration constant. Therefore functions as the one in (\ref{eqn_nueva}) admit an infinite number of vacuum dS solutions $H_0$. 

Let us now study the stability of de Sitter solutions in vacuum, where the perturbation is affected only by  the underlying gravitational theory. According to the equation for the perturbations (\ref{2.5}),  it yields,
\[
\left(16H_0^4f_{GG}^0+8H_0^2f_{RG}^0+f_{RR}^0\right)\ddot{\delta}(t)+\left(48H_0^5f_{GG}^0+24H_0^3f_{RG}^0+3H_0f_{RR}^0\right)\dot{\delta}(t)
\]
\be
-\left(64H_0^6f_{GG}^0-\frac{1}{3}f_{R}^0+32H_0^4f_{RG}^0+4H_0^2f_{RR}^0\right)\delta(t)=0\ .
\label{dS3}
\ee
Therefore, the stability of each dS solution depends on the values of the function $f(R,G)$ and its derivatives evaluated in $\{R_0, G_0\}$. The general solution of equation (\ref{dS3}) can be easily obtained, and  is given by,
 \begin{eqnarray}
 \delta(t)=C_1\e^{\mu_+t}+C_2\e^{\mu_-t}\ ,
 \label{solution_dS}
 \end{eqnarray}
 where
\begin{eqnarray}
  \mu_{\pm}\,=\,9H_0\l F^0\, \pm\sqrt{3F^{0}\left(-4f_{R}^0+75H_0^2 F^{0}\right)}.
 \label{dS4}
 \end{eqnarray} 
where the variable $F^{0}\equiv 16H_0^4f_{GG}^0+8H_0^2f_{RG}^0+f_{RR}^0$ was introduced to lighten the notation. 
%
The growth of the perturbation will depend both 
upon the overall sign of the parameters $\mu_{\pm}$ in expression (\ref{dS4}) and also upon the real or imaginary character of the square root. Thus, four different cases can be distinguished:

 
 \begin{itemize}

\item  $F_0<0$ and $f_R^0<75H_0^2F^0/4$: implies complex solutions and $\Re(\mu_{\pm})<0$, therefore solutions behave as a damped oscillator of decreasing amplitude. Solutions are thus stable.

\item $F_0>0$ and $f_R^0>75H_0^2F^0/4$: implies complex solutions and $\Re(\mu_{\pm})>0$, therefore solutions as an oscillator of increasing amplitude. Solutions are thus non-stable.

\item $F_0>0$ and $\frac{4f_R^0}{75H_0^2F^0}\in(0,1)$: implies that both $\mu_{\pm}$ are real and $\mu_{+}>0$. Solutions are thus non-stable.

\item $F_0<0$ and $\frac{4f_R^0}{75H_0^2F^0}\in(0,1)$: implies
that both $\mu_{\pm}$ are real and $\mu_{-}<0$. In this case, 
$\mu_{+}>0$ provided that  $|f_R^{0}|> |12 H_{0}^{2}F_0|$ 
and $\mu_{+}<0$ whenever $|f_R^{0}|< |12 H_{0}^{2}F_0|$.
Therefore the solutions are non-stable and stable respectively.
\end{itemize} 
 
According to the Appendix definitions, since $F^{0}=-\frac{c_2}{18 H_0^2}$, then $c_2>0$, i.e., 
the positivity of the second order coefficient in the perturbation equation (\ref{2.5}),
turns out to be the necessary (but not sufficient) condition to provide stability of the dS solutions in vacuum. 

In order to illustrate the previous calculations, let us consider the function
 \be
 f(R,G)=\kappa_1R+\kappa_2R^nG^m\ .
 \label{dS5}
 \ee
 Here $\{\kappa_1,\kappa_2\}$ are positive coupling constants. For simplicity we take $n=1$ and $m=3$, so that the equation (\ref{dS2}) can be solved, and we find the following dS solution,
 \be
 H_{0}=\frac{1}{2\sqrt{3}}\left(\frac{\kappa_1}{16\kappa_2}\right)^{1/6}\ .
 \label{dS6}
 \ee
Thus, considering the perturbation solution (\ref{dS4}), and introducing the function (\ref{dS5}) and its derivatives evaluated in (\ref{dS6}), the solution for the perturbation is determined, 
 \[
 \delta(t)\,=\,\left[C_1\exp\left(\frac{\sqrt{25\kappa_1\kappa_2-2^{23/3}3^{2}\kappa_1^{1/3}\kappa_2^{5/3}}}{\sqrt{3}\sqrt[3]{2^{5}\kappa_1\kappa_2^{2}}}t\right)
+ C_2\right] 
\]
\be 
\times \exp\left({-\frac{3\sqrt{\kappa_1}+\sqrt{25\kappa_1-2^{23/3}3^{2}\kappa_1^{1/7}\kappa_2^{2/3}}}{\sqrt{3}\sqrt[3]{2^{8}\kappa_1}\kappa_2^{1/6}}}t\right)
 \label{dS7}
 \ee
Hence, the stability depends on the value of the exponential of the first term in (\ref{dS7}), which is given by the values of the coupling constants $\kappa_1$ and $\kappa_2$. In this sense, for  $\kappa_1>2\left(\frac{1152}{25}\right)^{3/2}\kappa_2$, the perturbation grows exponentially, and the dS solution becomes unstable, otherwise the perturbation turns out to behave as a damped oscillator that tends to zero, so that the solution becomes stable.
 

\section{Stability of power-law solutions}	

In this Section we are interested in cosmological solutions of the type,
\be
a(t)\propto t^{\alpha} \rightarrow H(t)=\frac{\alpha}{t}\ \ .
\label{pw1}
\ee
that shall be referred to as power laws. These solutions represent the scale factor evolution 
for standard fluids, such as dust ($\alpha=2/3$) or radiation ($\alpha=1/2$) dominated Universe provided that GR is assumed as the underlying valid gravitational theory.

For the solutions (\ref{pw1}), the Gauss-Bonnet term and the Ricci scalar, given by expressions (\ref{1.5}), take respectively the following form,
\be
G=\frac{24 \alpha^3(\alpha-1)}{t^4}, \quad R=\frac{6\alpha(2\alpha-1)}{t^2}\ .
\label{pw4}
\ee

Let us now explore two different kinds of gravitational Lagrangian $f(R,G)$ able to mimic  the power-law solutions described by (\ref{pw1}) in the presence of standard fluids.  It is obvious that one has to
solve first the background FRLW equation (\ref{1.4_back}) in order to find the appropriate function $f(R,G)$.  The considered two types of $f(R,G)$ Lagrangians are the following

\subsection{$f(R,G)=f_1(G)+f_2(R)$}
For the sake of simplicity,  let us start by considering the subfamily of functions given in (\ref{2.9}). Then, for this type of Lagrangian the Friedmann equation (\ref{1.4_back}) can be split  into two equations \cite{arXiv:1001.3636}, as
\begin{eqnarray}
-24H^3\dot{G}\left(f_{1}\right)_{GG}+ G \left(f_{1}\right)_{G}-f_1\,&=&\,0\ , \nonumber\\
-3H\dot{R}\left(f_{2}\right)_{RR}+3(\dot{H}+H^2)\left(f_{2}\right)_{R}-\frac{1}{2}f_2 +\kappa^2\rho_{m0}\,&=&\,0
\label{pw3}
\end{eqnarray}
The first equation in (\ref{pw3}) can be written in terms of $G$, as
\begin{equation}
 G^2\left(f_{1}\right)_{GG}+\frac{\alpha-1}{4}G\left(f_{1}\right)_{G}-\frac{\alpha-1}{4}f_1=0\ ,
\label{pw5}
\end{equation}
which is an Euler equation, whose solution is easily obtained, and yields
\begin{equation}
f_1(G)=C_1 G^{\frac{1-\alpha}{4}}+C_2 G\ ,
\label{pw6}
\end{equation}
where $C_{1,2}$ are integration constants. Note that the second term in (\ref{pw6}) is the trivial Gauss-Bonnet solution and can be removed, since it becomes a 
total derivative and does not contribute to the field equations.  In the same way, the second equation in (\ref{pw3}) takes the form
\begin{equation}
R^2\left(f_{2}\right)_{RR}+\frac{\alpha-1}{2}R\left(f_{2}\right)_{R}-\frac{2\alpha-1}{2}f_2+\kappa^2(2\alpha-1)\rho_{m0}\,=\,0\ ,
\label{pw7}
\end{equation}
In the presence of a perfect fluid, whose equation of state is given by  $p_{m0}\,=\,w_m\rho_{m0}$, from the energy conservation equation (\ref{1.5}) with the class of cosmological solutions given in (\ref{pw1}),  the energy density yields,
\begin{equation}
\rho_{m0}=\rho_{m0}(t_{today}) t^{-3(w_m+1)\alpha}=\rho_{m0}(t_{today})\left[\frac{R}{6\alpha(2\alpha-1)} \right]^{\frac{3(1+w_m)\alpha}{2}}\ .
\label{pw9}
\end{equation}
where $\alpha\neq 1/2$  has been assumed\footnote{This is a special case to be discussed separately.}. Hence, the equation (\ref{pw7}) for $f_2(R)$ gives the solution,
\begin{eqnarray}
f_2(R)=k_1R^{\mu_+}+k_2R^{\mu_-}+\beta R^A\ ,  
\label{f2_with_matter}
\end{eqnarray}
where $k_{1,2}$ integration constants and
\[
\mu_{\pm}\,=\,\frac{3-\alpha(7-2\alpha)}{4(1-2\alpha)}\pm\frac{\sqrt{1+\alpha(10+\alpha)}}{4}\ , 
\]
\[
\beta\,\equiv \,\frac{-3\kappa^2\rho_{m0}(t_{today})(2\alpha-1)}{\left[6\alpha(2\alpha-1)\right]^{A}\left[A(A-1)+\frac{1}{2}A(\alpha-1)-\frac{2\alpha-1}{2}\right]}\ ,
\]
\be
A\equiv \frac{3(1+w_m)\alpha}{2}\ .
\label{6.3.8}
\ee
Then, the complete function $f(R,G)$ is reconstructed for this class of cosmological solutions, and can be expressed as the sum of the expressions 
(\ref{pw6}) and (\ref{f2_with_matter}).\\

Concerning the stability of this kind of solutions, we have to evaluate the function $f(R,G)$ and its derivatives in  (\ref{pw1}), and solve  the perturbation equations (\ref{2.5}) and (\ref{2.6}). 
However, analytical solutions for equation (\ref{2.5}) are in general difficult to be found. Only numerical solutions can be obtained by 
fixing the values of the free parameters of the theory, i.e., the coupling constants. 
In order to circumvent this intrinsic difficulty, let us consider the following particular cases: the radiation and dust matter evolutions, where $\alpha=2/3$ and $\alpha=1/2$ respectively, and $\alpha>1$, which gives an accelerating expansion.

\subsubsection{Dust dominated Universe, $\alpha=2/3$:}

\begin{figure*}[h!]
	\centering
		\includegraphics[width=0.4775\textwidth]{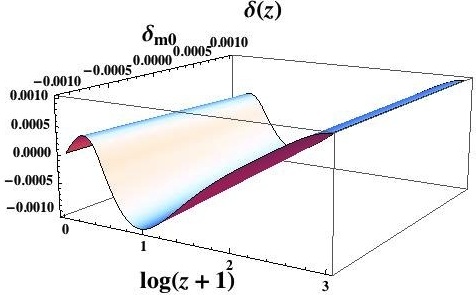}\,\,\,
		\includegraphics[width=0.4775\textwidth]{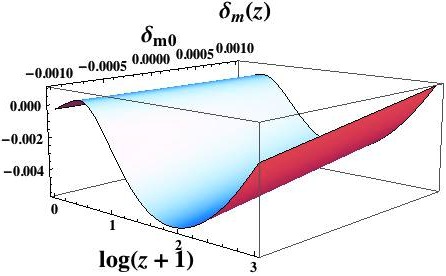}
		 \includegraphics[width=0.4775\textwidth]{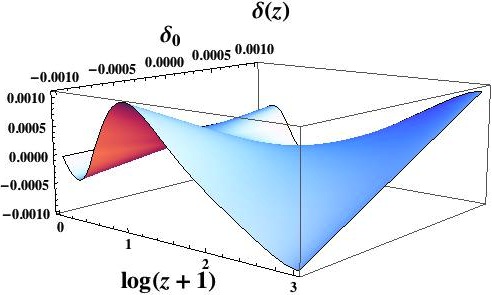}\,\,\
	\includegraphics[width=0.4775\textwidth]{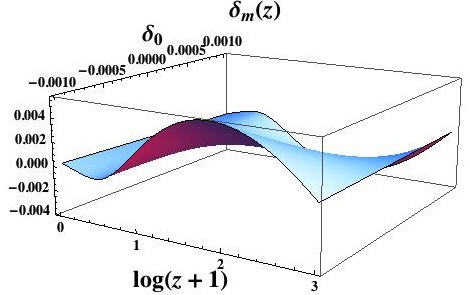}\,
 \caption{\footnotesize{
 The stability of the solutions for the model $f(R,G)=f_1(G)+f_2(R)$ is shown. Here we assume $\alpha=2/3$ and $w_m=0$, and the action reduces to $f(R,G)=f_2(R)$. The perturbations $\delta(z)$ and $\delta_{m}(z)$ are represented versus the redshift $z$. At the top figures, the initial condition $\delta_{m0}$ at $z=1000$ is varied while $\delta_0=0.001$ and $\delta'_0=0$ are fixed. At the bottom, $\delta_{m}(1000)=0.001$ is fixed while the evolution of $\delta(z)$ and $\delta_m(z)$ is shown as a function of $\delta_0$. As shown, independently of the initial conditions, the perturbations behave as a damped oscillator, turning out very small at $z=0$.}}
 \label{fig1}
\end{figure*}

\begin{figure*}[h!]
	\centering
		\includegraphics[width=0.4775\textwidth]{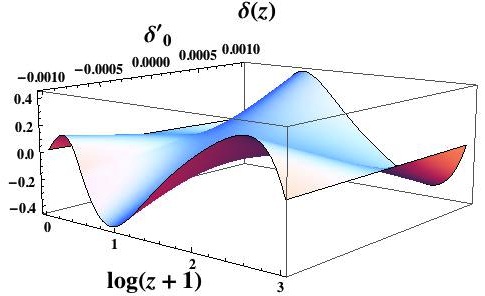}\,\,
		\includegraphics[width=0.4775\textwidth]{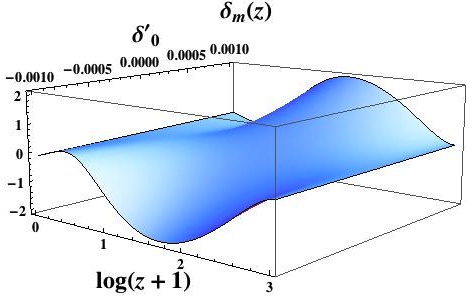}
 \caption{\footnotesize{
 As Fig.~\ref{fig1}, the stability for the solutions described by $\alpha=2/3$ and $w_m=0$ is analyzed. Here
 the perturbations $\delta(z)$ and $\delta_{m}(z)$ versus the redshift $z$ are represented and the initial conditions $\delta_0=\delta_{m0}=0.001$ are fixed, while $\delta'_0$ is varied along the range $\{-10^{-3},10^{-3}\}$.  Unlike Fig.~\ref{fig1}, the values of the initial condition $\delta'_0$ may produce a large amplitude in the oscillations of the perturbations along the evolution, leading to large effects at $z=0$.}}
 \label{fig2}
\end{figure*}

\begin{figure*}[h!]
	\centering
		\includegraphics[width=0.4775\textwidth]{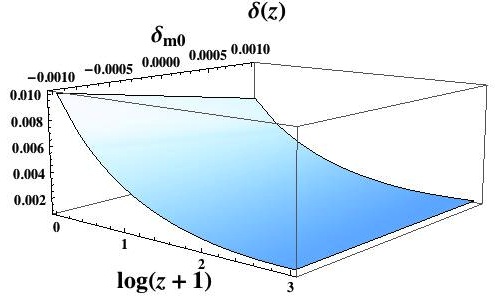}\,\,
		\includegraphics[width=0.4775\textwidth]{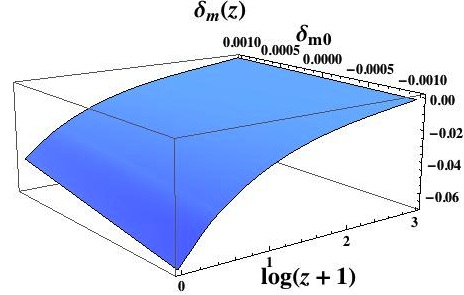}
		 \includegraphics[width=0.4775\textwidth]{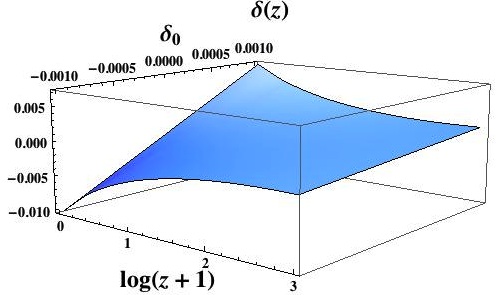}\,\,
	\includegraphics[width=0.4775\textwidth]{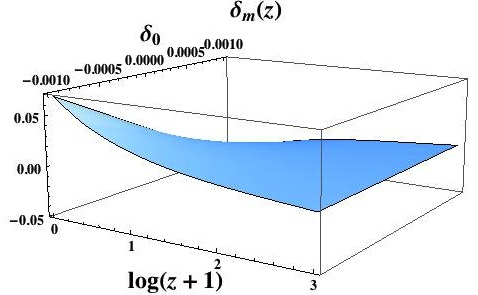}\,
 \caption{\footnotesize{
 The stability of the power law solution with $\alpha=2/3$ and $w_m=1/3$ is studied for the model (\ref{f2_with_matter}). As in Fig.~\ref{fig1}, at the top figures the initial condition $\delta_{m0}$ at $z=1000$ is varied whereas $\delta_0=10^{-3}$ and $\delta'_0=0$ are fixed. At the bottom, $\delta_{m}(1000)=10^{-3}$ is fixed while the evolution of $\delta(z)$ and $\delta_m(z)$ is shown as a function of $\delta_0$. Here, the perturbations for non null initial conditions  grow along the redshift evolution, producing instabilities at $z=0$.}}
 \label{fig3}
\end{figure*}
\begin{figure*}[h!]
	\centering
		\includegraphics[width=0.4775\textwidth]{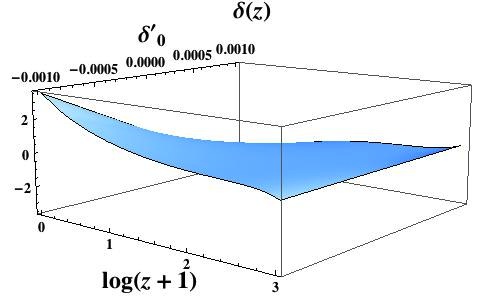}\,\,
		\includegraphics[width=0.4775\textwidth]{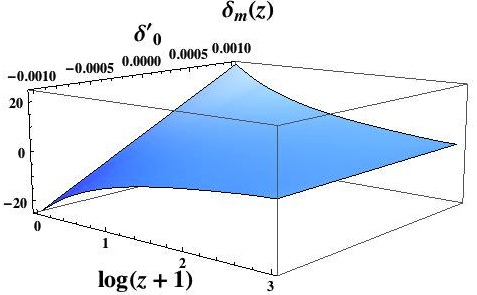}
 \caption{\footnotesize{
 As in Fig.~\ref{fig3}, the stability for the solutions given by $\alpha=2/3$ and $w_m=1/3$ is analyzed. However, here the perturbations $\delta(z)$ and $\delta_{m}(z)$ are represented versus the redshift $z$ with the initial conditions $\delta_0=\delta_{m0}=0.001$, whereas $\delta'_0$ is varied along the range $\{-10^{-3},10^{-3}\}$.  In this case, the perturbations grow very fast for non null initial conditions on $\delta'_0$, producing large instabilities at $z=0$.}}
 \label{fig4}
\end{figure*}
\begin{figure*}[h!]
	\centering
		\includegraphics[width=0.4775\textwidth]{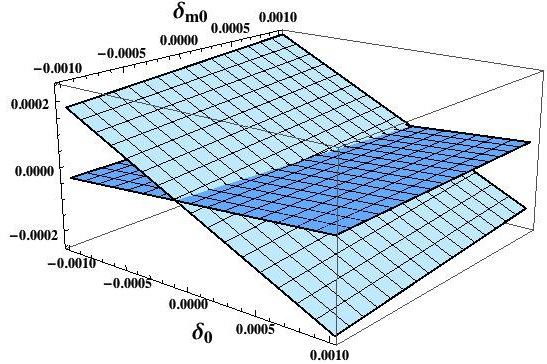}\,\,
		\includegraphics[width=0.4775\textwidth]{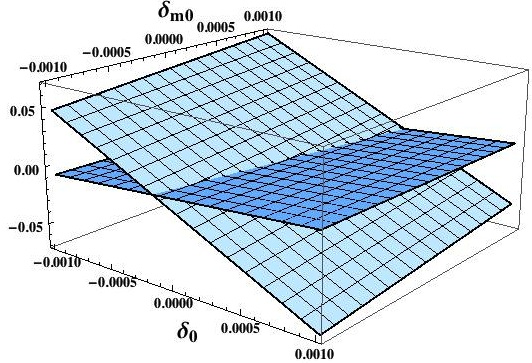}
 \caption{\footnotesize{
 This represents a summary of the stability of $f(R,G)=f_1(G)+f_2(R)$ for the power law solutions in the dust case $\alpha=2/3$: Evolution of  cosmological and dust perturbations $\delta$ and $\delta_m$ for $w_m=0$ (left panel) and $w_m=1/3$ (right panel) respectively.  Initial conditions $\{\delta_0,\delta_{m0}\}$ are imposed at redshift $z = 1000$ and varied in the range $\{-10^{-3},10^{-3}\}$, while $\delta'(z=1000)=0$ is fixed. In both figures, the hypersurfaces with larger slope correspond to $\delta_m(z=0)$, while the other hypersurfaces  corresponds to $\delta(z=0)$, which are not constant but with a much more little growth than $\delta_m$. In both cases, the evolution of the final values for the perturbations follow a linear relation with the initial conditions, as natural since the perturbations equations are studied in the linear regime.
}}
 \label{fig4a}
\end{figure*}

In the case of an expansion of the type of dust matter, given by $\alpha=2/3$, the Gauss-Bonnet term $G<0$ according to (\ref{pw4}), so that the gravitational Lagrangian (\ref{pw6}) becomes complex, which may be interpreted as a non-physical case. In order to avoid this scenario,  we set $C_1=0$, and thus the Lagrangian turns out $f(R,G)=f_2(R)$, given by (\ref{f2_with_matter}) .
%
For solving the differential equation (\ref{2.5}), we use numerical methods. In order to illustrate the richness of this case, we considered specific values for the coupling constants
$k_1=k_2=0.1\rho_{m0}$ with the appropriate dimensions. 
In figure \ref{fig1},  the evolution of $\delta(z)$ and $\delta_m(z)$ are plotted as functions of redshift $z$, and assuming different initial conditions for both perturbations. We also explore the effects of the variation of the initial condition for the first derivative $\delta'(z)$ in figure \ref{fig2}. We can see that regardless the initial conditions, the perturbation $\delta$ oscillates tending to zero at $z=0$ and similarly for $\delta_m$ in figures \ref{fig1} and \ref{fig2}. Moreover,  the case of a radiation-like fluid $w_m=1/3$  is explored in  figures \ref{fig3} and \ref{fig4}, where the perturbations increase at small values of the redshift $z=0$ in comparison with the pressureless case. While in figure \ref{fig4a}  the values $\delta(z=0)$ and $\delta_m(z=0)$ are plotted versus the initial conditions assumed at $z=1000$.


\subsubsection{Late-time acceleration, $\alpha>1$:}

\begin{figure*}[h!]
	\centering
		\includegraphics[width=0.4775\textwidth]{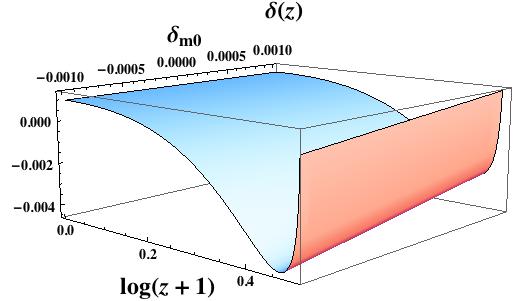}\,\,
		\includegraphics[width=0.4775\textwidth]{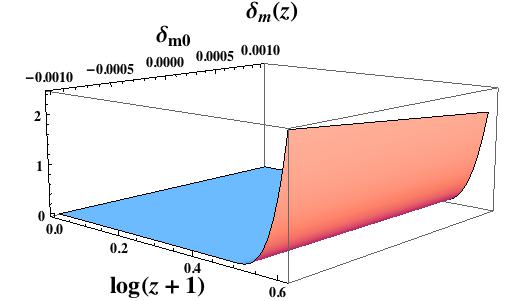}
		 \includegraphics[width=0.4775\textwidth]{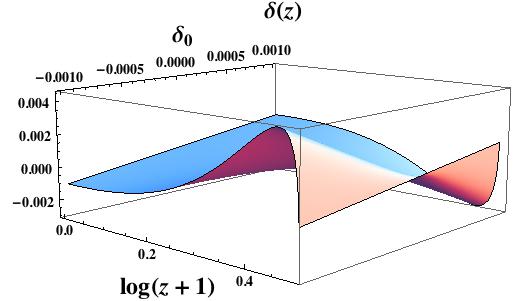}\,\,
	\includegraphics[width=0.4775\textwidth]{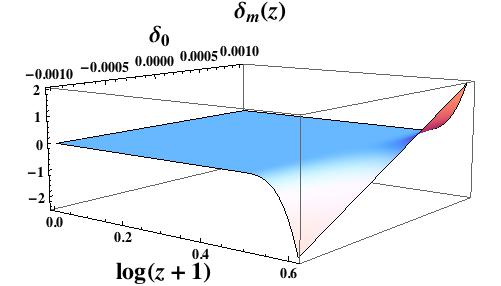}\,
 \caption{\footnotesize{Stability of $f(R,G)=f_1(G)+f_2(R)$ for $\alpha=2$: The evolution of the perturbations $\delta(z)$ and $\delta_{m}(z)$ versus the redshift $z$, and the initial values for $\delta_{m0}$ at the top, and $\delta_{0}$ at the bottom. Here we consider $\alpha=2$ and initial conditions imposed close to the pole $z\sim z_{pole}$ where  $\delta_{0}=0.001$ and $\delta_{0}'=0$ are fixed for the top figures, and $\delta_{m0}=0.001$ for the bottom figures. In both cases, we have assumed $w_m=0$. Different values of the coupling constants change the position of the pole $z_{pole}$. In all the cases the perturbation $\delta(z)$ behaves smoothly close to $z=0$ while it obviously becomes very large for $z$ close to the pole. In the same way $\delta_m(z)$ tends to small values close to $z=0$. }}
 \label{fig5}
\end{figure*}
\begin{figure*}[h!]
	\centering
		\includegraphics[width=0.7\textwidth]{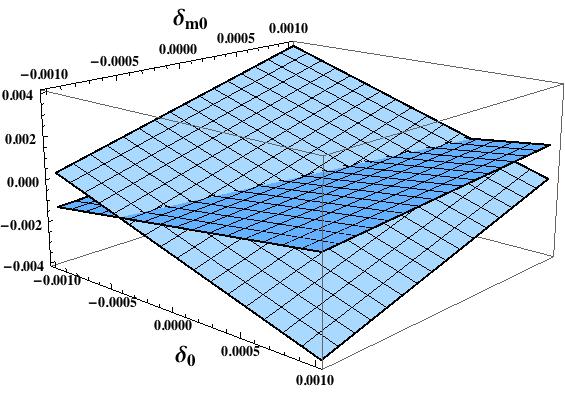}\,\,
 \caption{\footnotesize{
 Evolution of  cosmological and dust perturbations $\delta$ and $\delta_m$  for $\alpha=2$ with the presence of a pressureless fluid$w_m=0$.  Initial conditions $\{\delta_0,\delta_{m0}\}$ are imposed at redshift $z = 0$ and varied in the range $\{-10^{-3},10^{-3}\}$, while $\delta'(0)=0$ is fixed. As above, the hypersurface with larger slope corresponds to $\delta_m(z=2.08)$, while the other hypersurface  corresponds to $\delta(z=2.08)$. The final values are calculated at $z=2.08$, very close to the pole shown in fig.~\ref{fig5}. The perturbations do not become larger than the initial values assumed before the divergence. 
}}
 \label{fig6}
\end{figure*}
Another interesting example is given by a power-law cosmological solution with $\alpha>1$. 
In GR, this kind of solution is provided by the presence of a perfect fluid whose EoS parameter is given by $w<-1/3$ and consequently  compatible with the currently observed accelerating expansion.
Coming back to $f(R,G)$ gravity theories,  we are interested in studying its stability when  dust matter $w_m=0$ is included.  It can be shown that  the equation (\ref{2.5}) exhibits a pole, whose position in relation with the redshift depends upon the coupling constants values. For small redshifts the perturbation oscillates close to $z=0$, where the dark energy epoch is expected. Hence, it seems that in the presence of dust matter (baryons and cold dark matter), this kind of solutions becomes unstable for large redshifts, when the radiation/matter dominated epochs occur, and naturally the power-law with $\alpha>1$ is not valid anymore. However the solution tuns out to be stable when we approach to the current epoch, and late-time acceleration is well reproduced. For illustrative purposes,
 we have  considered $k_1=k_2=0.1\times\rho_0$ and $C_1\sim\rho_{m0}(t_{today})$ whose perturbations are depicted in Fig.~\ref{fig5}. In this case
the contribution coming from the Gauss-Bonnet part (\ref{pw6}) has to be included since $G>0$ and the action is therefore real. 
As in the cases above, different initial conditions for both $\delta$ and $\delta_{m}$ were considered, but now at $z=0$ since we are  interested in studying the past evolution of this solution with $\alpha>1$, as it  reproduces late-time acceleration at the current epoch. As shown in Fig.~\ref{fig5}, the perturbations diverge for large redshifts, so the solution is stable around $z=0$ but introduces large instabilities for large redshifts, where naturally the presence of a pressureless fluid (dust) should dominate. 
Moreover,  the values of $\delta(z=0)$ and $\delta_m(z=0)$ are evaluated depending on the initial conditions $\delta(z=2.08)$ and $\delta_m(z=2.08)$ in figure \ref{fig6}.

\subsubsection{Radiation dominated Universe, $\alpha=1/2$:}
In the case for $\alpha=1/2$, 
the equation (\ref{pw7}) has to be solved separately from the general case, which for $\alpha=1/2$, becomes, 
\be
R^2(f_2)_{RR}-\frac{1}{4}R(f_2)_R=0\ ,
\label{rad1a}
\ee
whose general solution is given by,
\be
f(R)=C_1R^{5/4}+C_2\ .
\label{rad1b}
\ee
Nevertheless, note that this is not the only solution for the case $\alpha=1/2$ since this value for the power-law exponent implies straightforwardly $R=0$. Thus, any function $f_2(R)$ that accomplish $\lim_{R\rightarrow0} f_2(R)\sim R^n$ with $n\geq0$, will satisfy the equation (\ref{rad1a}). However, by analyzing the 
Friedmann equation for $f_2(R)$ in (\ref{pw3}), in terms of the cosmic time instead of the Ricci scalar, one can show that no gravitational action other than Hilbert-Einstein ($n=1$) will satisfy this equation.
%
Let us prove the previous statement by considering an hypothetical action given by $f_2(R)\propto R^n$, the equation (\ref{pw3}) yields,
\be
6^n\alpha\frac{1-2\alpha+n(2n+\alpha-3)}{t^2}\left[\frac{\alpha(2\alpha-1)}{t^2}\right]^{n-1}+\kappa^2\rho_0 t^{-3(1+w_m)\alpha}=0
\label{rad1c}
\ee
Then, if $\alpha=1/2$ is assumed, the first term in the l.h.s. of (\ref{rad1c}) vanishes, whilst the second is never zero unless $\rho_{m0}=0$ (vacuum). Hence, the only possible solution is given for $n=1$, where the $f_2(R)$ action becomes the Hilbert-Einstein action, which imposes that $w_m=1/3$ as natural for a radiation type expansion. \\ 

Hence for $\alpha=1/2$, the gravitational action needs to be either $f(R,G)=R+f_1(G)$, or $f(R,G)\equiv f(G)$. In the first case, for $f_1(G)$  given by expression (\ref{pw6}), but since $G<0$ according to (\ref{pw4}), the action (\ref{pw6}) would become complex, and non physical, such that $f(R,G)=R$. For the second possibility, $f(R,G)\equiv f(G)$, the action only depends on the Gauss-Bonnet term, and the FLRW equations (\ref{pw3}) can be rewritten as,
\begin{equation}
 G^2f_{GG}+\frac{\alpha-1}{4}Gf_{G}-\frac{\alpha-1}{4}f+\kappa^2\frac{\alpha-1}{4}\rho_m=0\ ,
\label{rad1}
\end{equation}
while the expression for the energy density of a perfect fluid $p_m=w_m\rho_m$ is given by,
\be
\rho_{m}(t)\,=\,\rho_{m0}(t_{today})\,t^{-\frac{3(1+w_m)}{2}}=\rho_{m0}\left[\frac{G}{24\alpha^3(\alpha-1)}\right]^{\frac{3(1+w_m)\alpha}{4}}\ .
\label{rad2}
\ee
 Hence, the equation (\ref{rad1}) can be easily solved,
 \be
 f(G)=C_1 G^{\frac{1-\alpha}{4}}+C_2 G+\beta G^{A}\ ,
\label{rad3}
\ee
where
\[
\beta\,=\,\frac{\kappa^2\rho_{m0}(t_{today})}{4}\frac{1-\alpha}{\left[24\alpha^3(\alpha-1)\right]^A\left[(A-1)A+\frac{\alpha-1}{4}(A-1)\right]},
\]
\be
A\,=\,\frac{3(1+w_m)\alpha}{4}\,.
\label{rad4}
\ee
However, as above the Gauss-Bonnet term $G <0$ since $\alpha<1$  and  the gravitational action becomes complex, which in principle lacks any physical sense. 
Hence, in the case of $\alpha=1/2$, the gravitational action reduces to the Hilbert-Einstein action with the presence of a radiation-like fluid. This analysis provides an interesting consequence, since the universe evolution crosses a radiation dominated epoch, if the gravitational action is of the type of $f(R,G)=f_1(G)+f_2(R)$, the extra terms in the action should be negligible during such epoch, and the action must approach Hilbert-Einstein action.

\subsection{$f(R,G)\,=\,\mu R^{\beta}G^{\gamma}$}
Let us assume now a gravitational Lagrangian of the form
\begin{eqnarray}
f(R,G) = \mu R^{\beta}G^{\gamma}
\label{product_power_laws}
\end{eqnarray}
where $\mu$ is a dimensionful constant and $\beta$ and $\gamma$ are constant exponents an let us study the power-law solutions as the ones given at (\ref{pw1}) for two standard fluids, radiation an dust matter.

\subsubsection{Radiation dominated era, $\alpha=1/2$:}
For $f(R,G)$ models as given in expression (\ref{product_power_laws}), it is straightforward 
to prove that in the absence of any fluid and provided that $\beta>1$ and any even value of $\gamma$ (since $G<0$, this constraint gets rid of imaginary terms), they can mimic 
radiation-like scale factor evolution, i.e. $\alpha=1/2$. This can be performed by solving equation (\ref{1.4_back}) in the absence of 
fluid sources and considering a function like (\ref{product_power_laws}). For this kind of models, equation (\ref{2.5}) becomes

\begin{eqnarray}
t^2 \delta''(t)-\frac{1}{2}t \left(4\beta+8\gamma-9\right)\delta'(t)-\frac{1}{2}\left(4\beta+8\gamma-5\right)\delta(t)\,=\,0
\label{eqn_perturbations_Case_rad}
\end{eqnarray}
This is an Euler-like equation that yields to the solutions
\begin{eqnarray}
\delta(t)\,=\,C_+\,t^{\alpha_{+}}+C_{-} t^{\alpha_{-}}
\label{Solution_radiation}
\end{eqnarray}
where $C_{\pm}$ are arbitrary constants and 
\begin{eqnarray}
\alpha_{\pm}=\beta+\frac{1}{4}\left(-7+8\gamma\pm|8\gamma+4\beta-3|\right)
\end{eqnarray}
In order to understand the stability of the perturbation given by expression (\ref{Solution_radiation}), one should study the sign of $\alpha_{\pm}$ exponents: provided that the requirement $\beta>1$  is satisfied, it can be proved that $\alpha_{-}$ is always negative. Concerning $\alpha_{+}$, this exponent is negative provided that  $\gamma < 1/8 (5 - 4 \beta)$ and $\gamma$ even number as explained above. Otherwise $\alpha_+$ would be positive and the perturbation unstable. 

If one now considers a radiation fluid characterized by both $\omega=1/3$  and $\alpha=1/2$ it is possible to show that Lagrangians such as (\ref{product_power_laws}) cannot satisfy equation (\ref{1.4_back}) for any value of their parameters.  
 If we relax now the constraint on $\alpha$ permitting $\alpha\neq 1/2$, then the combination $\beta + 2\gamma$ has to be negative in order to guarantee $\alpha>0$ . It can be shown that in this case, equation (\ref{1.4_back}) is not satisfied for any parameter combination\footnote{In fact, the required $\mu$ would be imaginary.}.
Therefore, none of these two last cases can be accomplished by Lagrangians such as  (\ref{product_power_laws}).


\subsubsection{Dust dominated era, $\alpha=2/3$:}
\begin{figure*}[h!]
	\centering
		\includegraphics[width=0.4775\textwidth]{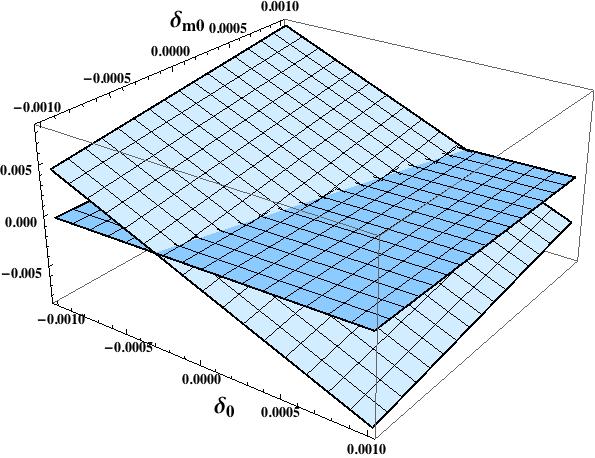} \,\,\,\,\,
		\includegraphics[width=0.4775\textwidth]{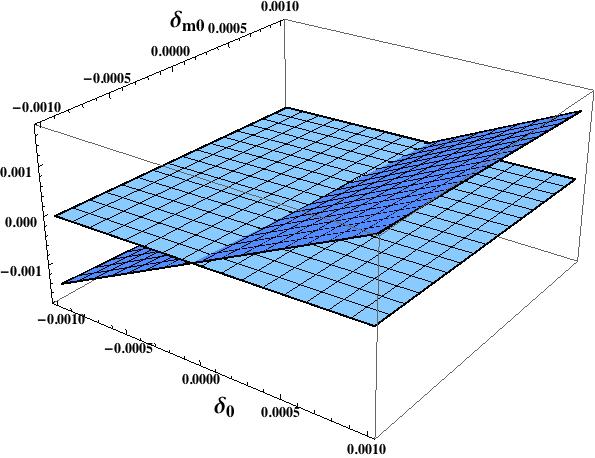}
%
		\caption{\footnotesize{
Values today $(z=0)$ for 
cosmological and matter (dust) perturbations, $\delta$ and $\delta_m$ respectively for $\beta=5$ (left panel) and $\beta=-1$ (right panel) in expression 
(\ref{eqn_perturbations_case2}) and $H_{0}(t)=2/3t$ in (\ref{2.6}). 
Initial conditions were imposed at redshift $z=1000$ ranging in the interval $(-0.001,\,+0.001)$ for both $\delta(z=1000)$ and $\delta_m(z=1000)$.
In both panels, the hyperplanes with larger slope correspond to $\delta_m$ evaluated today. The other hyperplanes which seem of constant value in the 3-dimensional representation correspond to $\delta$ today.
In both panels can be seen how $\delta_m$ 
achieves today values bigger -- in absolute value -- than the initial conditions. 
On the other hand, $\delta$ remains with amplitudes today smaller than the initial values.
%
%
		}}
	\label{fig:522_Valores_today}
\end{figure*}

\begin{figure*}[h!]
	\centering
		\includegraphics[width=0.4775\textwidth]{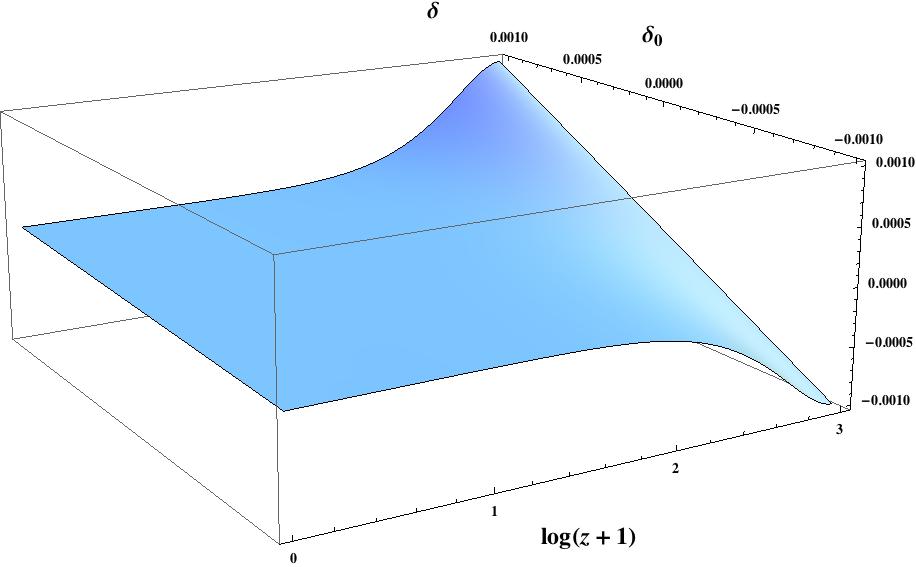}\,\,\,\,\,\,\,\,
		\includegraphics[width=0.4775\textwidth]{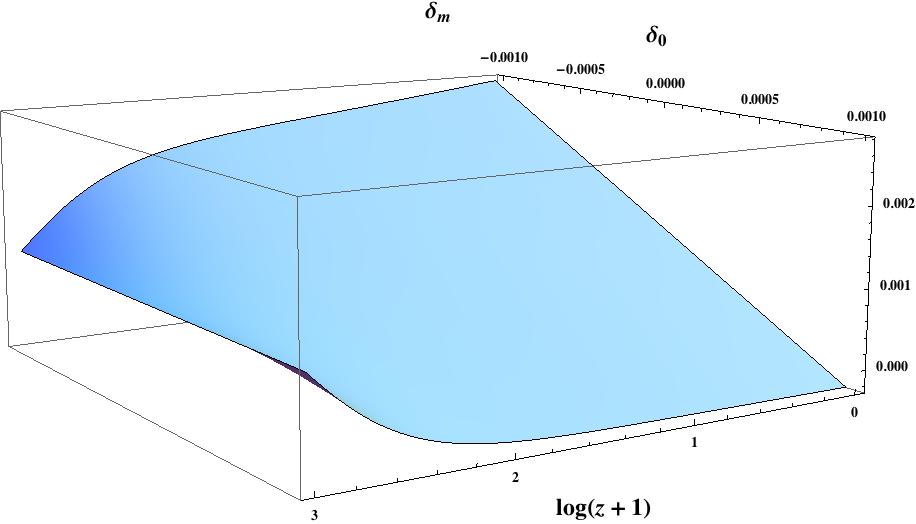}
		 \includegraphics[width=0.4775\textwidth]{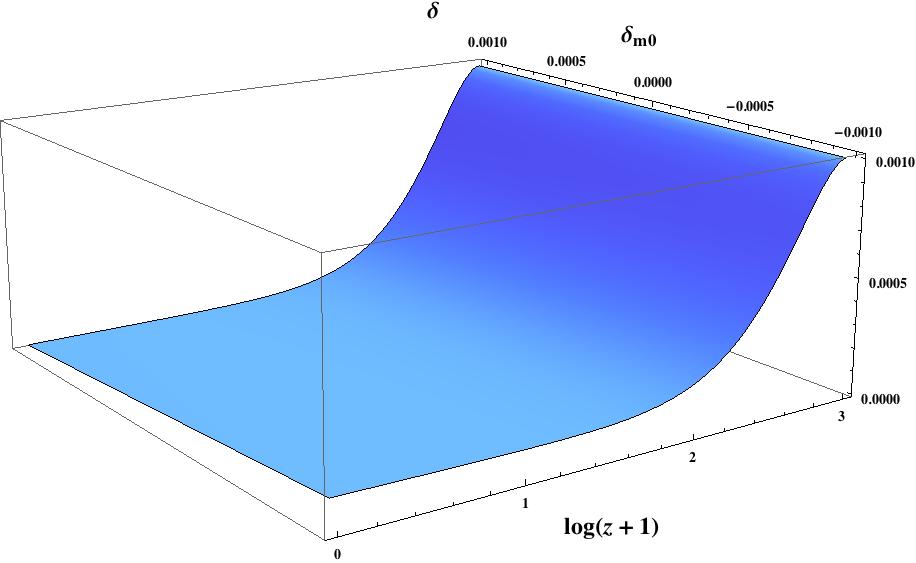}\,\,\,\,\,\,\,\,
		\includegraphics[width=0.4775\textwidth]{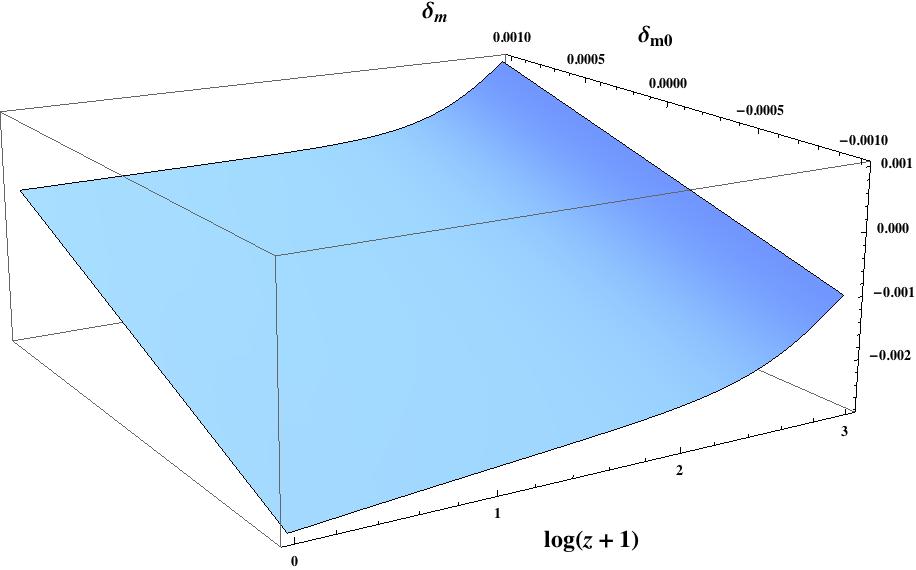}
		%
%
		\caption{\footnotesize{
			Stability of $f(R,G)$ given by (\ref{product_power_laws})  in the dust ($\omega=0$) case:
Evolution of cosmological and dust perturbations, $\delta$ and $\delta_m$ respectively for $\beta=5$ in expression 
(\ref{eqn_perturbations_case2}). 
Different initial conditions at redshift $z=1000$ were imposed. 
The two upper panels represent the evolution in redshift for $\delta$ (left) and $\delta_m$ (right) for 
fixed $\delta_m(z=1000)=10^{-3}$  whereas $\delta(z=1000)$ varied from $-0.001$ to $0.001$. One can see how the evolutions 
of both perturbations acquire decreasing amplitudes when approaching today.
At the lower figures, it is depicted again $\delta$ (left) and $\delta_m$ (right) this time for 
fixed $\delta(z=1000)=10^{-3}$  whereas $\delta_{m}(z=1000)$ varied from $-0.001$ to $0.001$. On the left panel it can be seen how $\delta$ tends to null amplitude today regardless
the values for $	\delta_m(z=1000)$. On the contrary, the values of $\delta_m$ (right panel)
 tend to decrease in amplitude by depending on its initial value.
		}}
	\label{fig:522_ValoresRedshift_beta_5}
\end{figure*}

\begin{figure*}[h!]
	\centering
		\includegraphics[width=0.4775\textwidth]{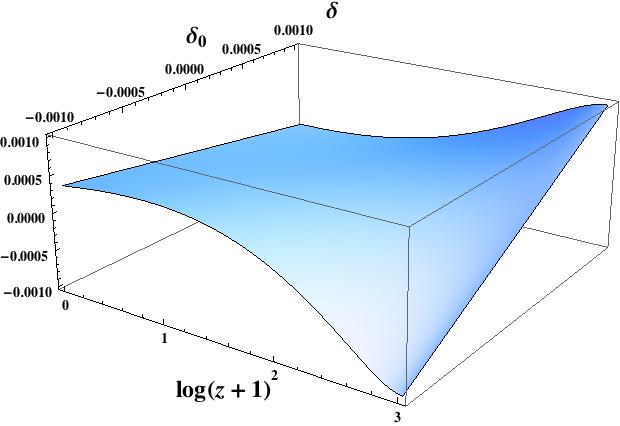}\,\,\,\,\,\,\,\,
		\includegraphics[width=0.4775\textwidth]{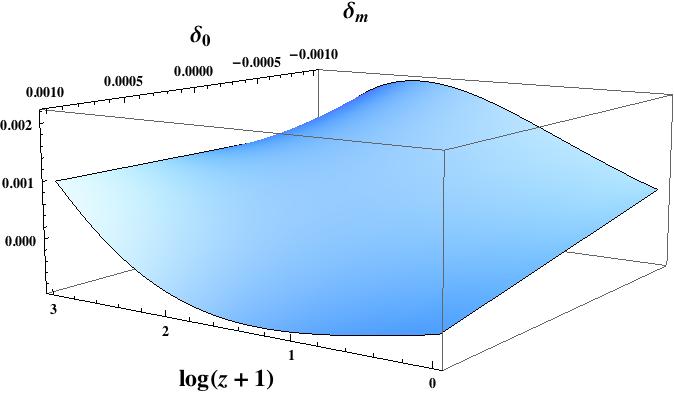}
		 \includegraphics[width=0.4775\textwidth]{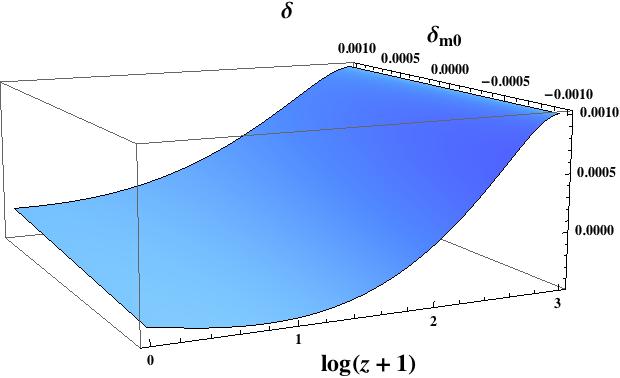}\,\,\,\,\,\,\,\,
		\includegraphics[width=0.4775\textwidth]{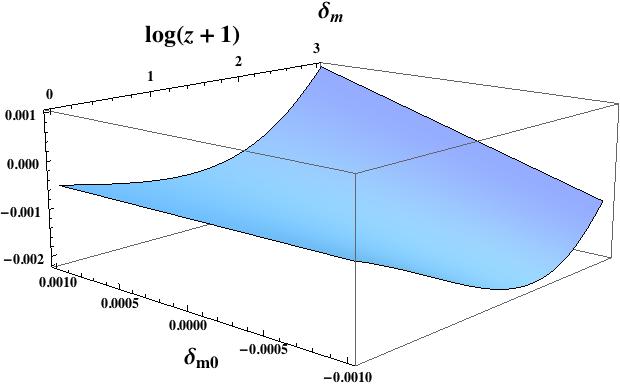}		
%
%
		\caption{\footnotesize{
			Stability of $f(R,G)$ given by (\ref{product_power_laws})  in the dust ($\omega=0$) case:
Evolution of cosmological and dust perturbations, $\delta$ and $\delta_m$ respectively for $\beta=-1$ in expression 
(\ref{eqn_perturbations_case2}). 
Different initial conditions at redshift $z=1000$ were imposed. 
The two upper panels represent the evolution in redshift for $\delta$ (left) and $\delta_m$ (right) for 
fixed $\delta_m(z=1000)=10^{-3}$  whereas $\delta(z=1000)$ varied from $-0.001$ to $0.001$. One can see how the $\delta$ 
perturbations tend to decrease its amplitudes when approaching today. Concerning $\delta_m$, with initial amplitude of $10^{-3}$, acquired final 
amplitudes ranging from $0$ to $10^{-3}$. 
At the lower figures, it is depicted again $\delta$ (left) and $\delta_m$ (right) this time for 
fixed $\delta(z=1000)=10^{-3}$  whereas $\delta_{m}(z=1000)$ varied from $-0.001$ to $0.001$. 
On the left panel it can be seen that $\delta$ tends today to amplitudes ranging between zero and $-5\cdot10^{-4}$, i.e., the amplitudes decrease 
for all the $\delta_m$  initial conditions. Concerning the evolution for $\delta_m$ (right panel), the amplitudes today are smaller than the corresponding initial amplitudes.
		}}
	\label{fig:522_ValoresRedshift_beta_minus1}
\end{figure*}

For this case, we have considered three different scenarios:
In the absence of matter, models given by (\ref{product_power_laws}) can hold a power-law scale factor with $\alpha=2/3$. In this case, 
$\mu$ can take in principle any value whereas $\gamma$ (which has to be even as in the previous case) and $\beta$ must be related as follows
\begin{eqnarray}
\gamma\,\equiv\,\gamma_{\pm} \,=\,\frac{1}{24}\left(13+6 \beta \pm \sqrt{121-180\beta+324 \beta^2}\right)
\label{gammas}
\end{eqnarray}
in order to satisfy the background equation (\ref{1.4_back}). With this requirement, the perturbation equation (\ref{2.5}) becomes a Euler-like equation

\[
t^2\ddot{\delta}(t)-\frac{1}{6}\left[-17+18\beta\pm\sqrt{121+36\beta(-5+9\beta)}\right] t \dot{\delta}(t)
\]%
\be
+\frac{1}{6}\left[
5-18\beta\mp\sqrt{121+36\beta(-5+9\beta)}
\right]\delta(t)=0
 \label{case1_dust}
 \ee
whose solutions are
\begin{eqnarray}
\delta_{\pm}(t)\,=\,\frac{C_1}{t}+C_2 t^{-3 + 2 \beta + 4 \gamma_{\pm}}
\label{solution_matter_in_vacuum}
\end{eqnarray}
where the second exponent is always a real number regardless the $\beta$ value. Moreover, it can be proved that for $\delta_{+}(t)$", i.e. $\gamma_{+}$ choice, the second power-law in (\ref{solution_matter_in_vacuum}) possesses a positive exponent, whereas $\delta_{-}(t)$ holds both power-law solutions with negative exponents. Therefore, the solutions provided by the $\gamma_{-}$ choice would be stable whereas the ones by $\gamma_{+}$ would not be.\\

%
The second case consists in considering the presence of dust matter ($\omega=0$) and a power-law scale factor 
with exponent $\alpha=2/3$. In that case $\gamma$ and $\mu$ depend on $\beta$ as follows
\begin{eqnarray}
\gamma\,=\,\frac{1-\beta}{2}\;\;,\;\;
\mu \, = \,\frac{(2\imath)^{\beta-1}3^{(3-\beta)/2}\kappa^2\rho_{m0}(t_{today})}{9\beta-5}.
\label{restrictions1_case2}
\end{eqnarray}
From the last expression, it is straightforward to conclude that one of the two following constraints must be imposed to guarantee $\mu$ to be positive and real:
\begin{eqnarray}
\beta\,=\, 4n+1 \,\,\,and\,\,\, \beta>\frac{5}{9} \;\;\;;\;\;\;\;
\beta\,=\,4n+3 \,\,\,\ and\,\,\,  \beta<\frac{5}{9} \;\;\;  with \,\,\,n\in \mathbb{Z} 
\label{restrictions2_case2}
\end{eqnarray}
For this case, the equation (\ref{2.5}) is again Euler-type with a source term proportional to $\delta_M(t)$. Thus
%
\[
t^2\ddot{\delta}(t)+ 3 t \dot{\delta}(t)+\left[1+\frac{8}{45(-1+\beta)}+\frac{12}{5+45\beta}\right]\delta(t)\,=\,\frac{2(-5 + 9 \beta)}{9[1+(8-9 \beta)\beta]}\delta_{M}(t) 
\]
\bea
\label{eqn_perturbations_case2}
\eea
%


In order to illustrate the rich phenomenology of this case, we have considered $\beta=5$ for the first 
constraint\footnote{Note that $\beta=1$ will imply $\gamma=0$  , i.e., the usual EH Lagrangian would be recovered.} 
and $\beta=-1$ for the second according to (\ref{restrictions2_case2}). For these two cases, in Figure \ref{fig:522_Valores_today} we have plotted the values for $\delta$ and 
$\delta_m$ today for a wide range  of initial conditions.
On the one hand, for $\beta=5$, the homogeneous Euler-like associated equation to (\ref{eqn_perturbations_case2}) would present complex 
conjugate exponents. 
%
%
The presence of the matter term induces solutions as the ones presented in Figures 
\ref{fig:522_ValoresRedshift_beta_5}. 
In this figure, the perturbations amplitudes for $\delta$ seem to decrease whereas
those for $\delta_m$ increase but remaining all of them in the linear regime. 
%
On the other hand, for $\beta=-1$ the homogeneous associated equation to (\ref{eqn_perturbations_case2}) 
would present two distinct negative real exponents.  
The inclusion of the matter term leads to the perturbations amplitudes to decrease in absolute value 
as presented in Figure \ref{fig:522_ValoresRedshift_beta_minus1} both for matter and Hubble parameter perturbations.
%
%

\begin{figure*}[h!]
	\centering
		\includegraphics[width=0.7\textwidth]{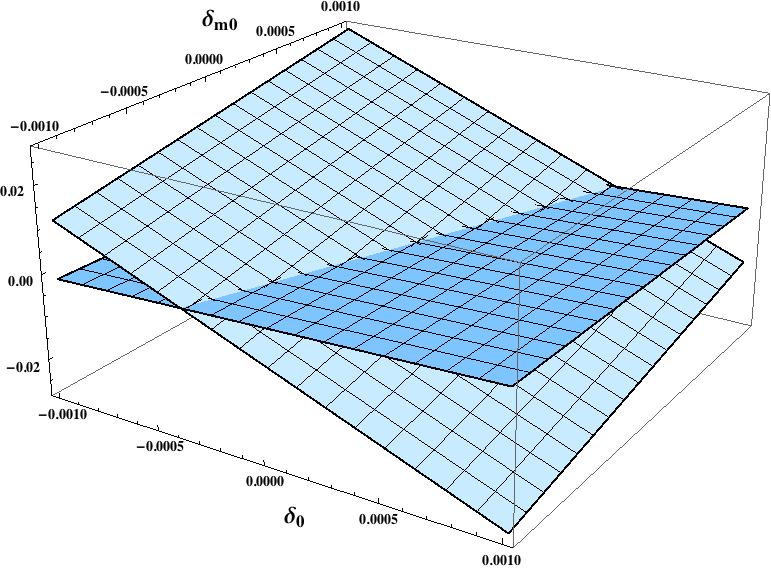} 
%
		\caption{\footnotesize{
Values today for 
cosmological and matter (dust) perturbations, $\delta$ and $\delta_m$ respectively, for the power-law scale factor with exponent 
$\alpha=4/3$:
Parameters of the gravitational Lagrangian were chosen to be
$\beta=-1$,  $\gamma = 3/2$ and $\mu=75/256 \sqrt{3/2} \kappa^2\rho_{m}(t_{today})$.
The hyperplane with larger slope corresponds to $\delta_m$ evaluated today whose values
range from $-0.03$ to $0.03$, i.e., $30$ times bigger than initial amplitudes and therefore showing the instability of the $\delta_m$ evolution. 
With regard to $\delta$, the hyperplane for $\delta$ today acquires amplitudes ranging from $-3\cdot10^{-3}$ to $3\cdot10^{-3}$, i.e., three times the initial value and consequently showing as well 
the instability of the $\delta$ evolution. 
		}}
	\label{fig:523_Valores_today}
\end{figure*}
\begin{figure*}[h!]
	\centering
		\includegraphics[width=0.4775\textwidth]{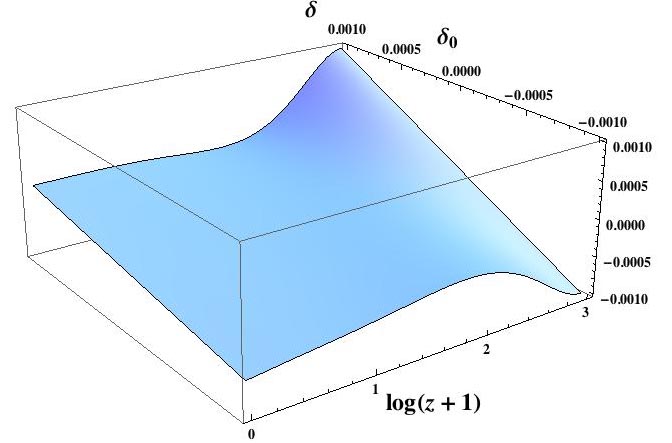}\,\,\,\,\,\,\,\,
		\includegraphics[width=0.4775\textwidth]{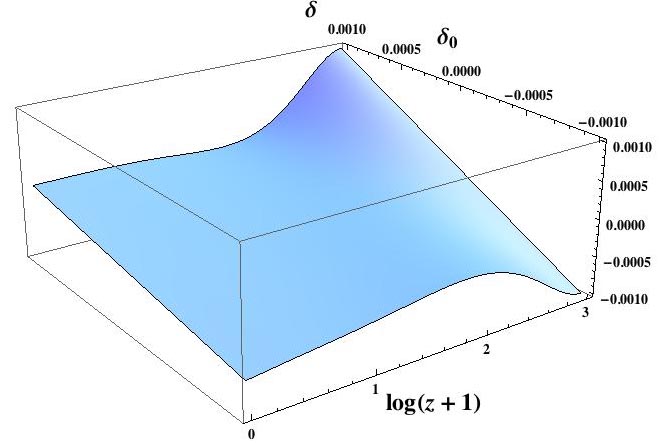}
		 \includegraphics[width=0.4775\textwidth]{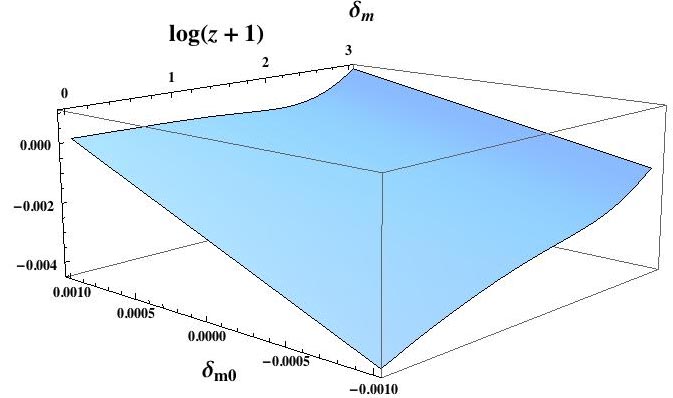}\,\,\,\,\,\,\,\,
		\includegraphics[width=0.4775\textwidth]{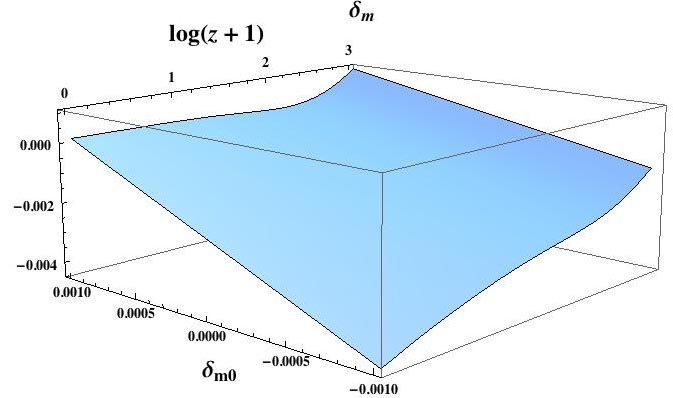}		
%
		\caption{\footnotesize{
					Stability of $f(R,G)$ given by (\ref{product_power_laws})  in the case of dust ($\omega=0$) and scale factor with power-law exponent 
					$\alpha=4/3$:
Evolution in redshift of cosmological and dust perturbations, $\delta$ and $\delta_m$ respectively.
As in Figure \ref{fig:523_Valores_today}, parameters of the gravitational Lagrangian were chosen to be
$\beta=-1$,  $\gamma = 3/2$ and $\mu=75/256 \sqrt{3/2} \kappa^2\rho_{m}(t_{today})$.
Different initial conditions at redshift $z=1000$ were imposed. 
The two upper panels represent the evolution in redshift for $\delta$ (left) and $\delta_m$ (right) for 
fixed $\delta_m(z=1000)=10^{-3}$  whereas $\delta(z=1000)$ varied from $-0.001$ to $0.001$. One can see how the $\delta$ 
perturbations tend to reduce its amplitudes when approaching today showing the stability of this case. Concerning $\delta_m$, with initial amplitude of $10^{-3}$, acquired final 
amplitudes ranging from $10^{-3}$ to $4\cdot 10^{-3}$,i.e., increasing its amplitude by a factor $4$. 
At the lower figures, it is depicted again $\delta$ (left) and $\delta_m$ (right) this time for 
fixed $\delta(z=1000)=10^{-3}$  whereas $\delta_{m}(z=1000)$ varied from $-0.001$ to $0.001$. On the left panel it can be seen how $\delta$ today acquires values ranging from 
$0$ to $5\cdot10^{-4}$, i.e., decreasing its amplitude (initially $10^{-3}$) for the $\delta_m$ initial condition. Concerning the evolution for $\delta_m$ (right panel), the amplitudes today range from $0$ (for $\delta_m(z=1000)=10^{-3})$ to $-4\cdot10^{-2}$ (for $\delta_m(z=1000)=-10^{-3})$.
		}}
	\label{fig:523_ValoresRedshift}
\end{figure*}

Finally, we decided to study a dust fluid characterized by both $\omega=0$  but 
$\alpha\neq2/3$. It is possible to show that Lagrangians such as (\ref{product_power_laws}) can satisfy equation (\ref{1.4_back}) provided that 
\begin{eqnarray}
\alpha\,=\,\frac{2}{3}\left(\beta +2\gamma\right)\,,
\label{Case_III_p}
\end{eqnarray}
and $\mu$ a certain combination of $\{\beta, \gamma, \rho_{m}(t_{today})\}$ that guarantees this parameter to be both real and positive. 
Therefore, models like (\ref{product_power_laws})  may host dust fluid ($\omega=0$) with a power-law evolution exponent $\alpha\neq2/3$ for a suitable choice of parameters.
In order to illustrate this case, we consider one example with $\beta=-1$ and  $\gamma = 3/2$ and consequently $\mu=75/256 \sqrt{3/2} \kappa^2\rho_{m}(t_{today})$. According to expression (\ref{Case_III_p}), the obtained value for the $\alpha$ exponent is $4/3$. 
%
%
%
Figure \ref{fig:523_Valores_today} represents $\delta$ and $\delta_m$ evaluated today for a wide range of initial conditions for these two quantities. On the other hand, Figure 
\ref{fig:523_ValoresRedshift} represents the redshift evolution also for $\delta$ and $\delta_m$ by fixing different initial conditions.
There, it can be seen how whereas the Hubble parameter perturbation remains small and decreases in amplitude, the matter 
perturbation grows in amplitude whereas remaining in the linear regime.

\section{Stability of $f(R,G)$ mimicking $\Lambda$CDM solution}

We analize in this section the stability of the $f(R,G)$ function mimicking $\Lambda$CDM 
cosmological evolution without any cosmological constant term. This model was originally presented in Following \cite{arXiv:1001.3636} and its key features 
will be described below. The cosmological speed-up effect of the cosmological constant in the 
Concordance model is precisely replaced by the modification introduced by $f(R,G)$ with respect to the usual EH Lagrangian.

The scale factor solution within GR with dust and cosmological constant is given by:
\begin{eqnarray}
a(t)\,=\,\left(\frac{\Omega_m}{\Omega_{\Lambda}}\right)^{2/3}\sinh^{1/3}\left(\frac{3\sqrt{\Omega_{\Lambda}}}{2}H_0\,t\right)
\label{a_LCDM}
\end{eqnarray}
\\
where $H_0$ holds for Hubble parameter today and $\Omega_{m, \Lambda}$ hold respectively for dust and cosmological constant 
fractional densities today\footnote{For illustrative purposes we shall consider $\Omega_m=0.27$ and $\Omega_{\Lambda}=0.73$.}. 
According to \cite{arXiv:1001.3636}, the Gauss-Bonnet contribution to the gravitational action that is able to mimic $\Lambda$CDM evolution is given by 
\begin{eqnarray}
 f(G)=\theta\ \zeta(G)^2 + \vartheta\ \zeta(G) + H_{0}^2 \epsilon\ ,
\end{eqnarray}
where
\begin{eqnarray}
\zeta(G)=\frac{3H_0^2\pm\sqrt{81H_0^4-3G}}{6l}\ ,
\label{GBLCDM}
\end{eqnarray}
with $l=\kappa^2\rho_0 a(t=t_{today})^{-3}/3$ 
and $G$ is given by expression (\ref{1.4}), while according to \cite{arXiv:1001.3636}, $\{\theta,\vartheta\}$ are
\begin{eqnarray}
\theta\,=\, \frac{l^2}{H_0^2}\left[\kappa^{-2}-\frac{2}{9}(\epsilon+1)\right]\;\;;\;\;\vartheta\,=\, l\left[\frac{1}{5}\kappa^{-2}+\frac{2}{9}(\epsilon+1)\right]\ ,
\end{eqnarray}
where $\epsilon$ is a constant\footnote{In the original reference, authors used $\delta$ symbol for this constant. In order to avoid confusion, we have preferred to use $\epsilon$ symbol.} as well. Therefore,  the full gravitational Lagrangian is expressed as follows \cite{arXiv:1001.3636}
\begin{eqnarray}
f(R,G)\,=\,R+\frac{1}{2}\left(\theta\ \zeta(G)^2 + \vartheta\ \zeta(G) + H_{0}^2 \epsilon \right)
\label{model_LCDM}
\end{eqnarray}

\begin{figure*}[h!]
	\centering
		\includegraphics[width=0.4775\textwidth]{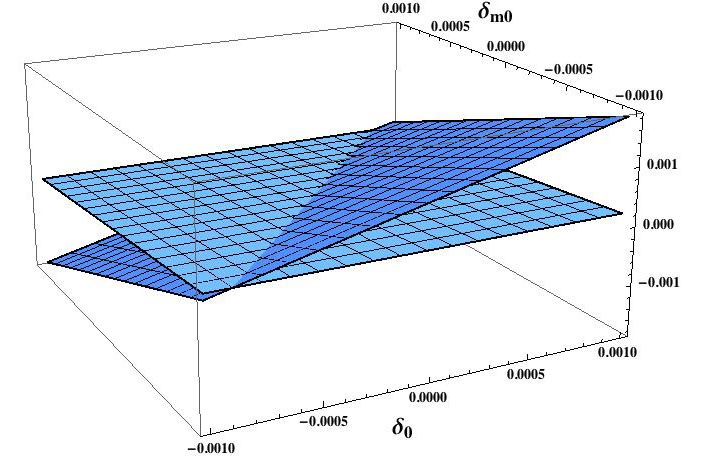}\,\,\,\,\,
		\includegraphics[width=0.4775\textwidth]{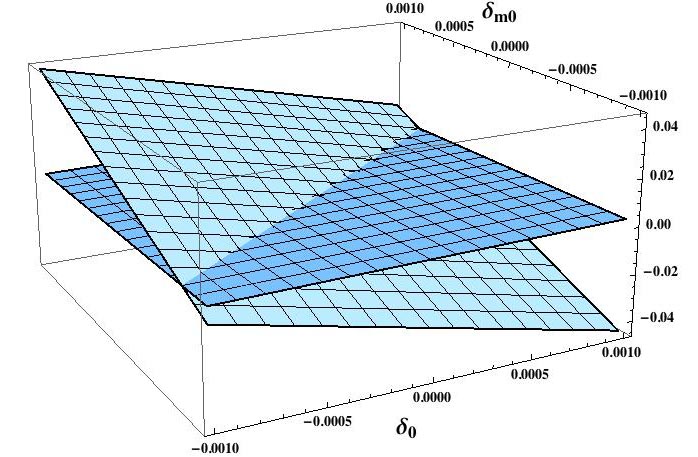}
%
		\caption{\footnotesize{Values today for cosmological and dust perturbations, $\delta$ and $\delta_m$ respectively, for the $f(R,G)$ model (\ref{model_LCDM}) mimicking $\Lambda$CDM. Positive (negative) sign in expression (\ref{GBLCDM}) were represented in left (right) panels. Initial conditions were imposed at redshift $z=1000$ ranking from $(-0.001,\,+0.001)$ for both $\delta(z=1000)$  and $\delta_m(z=1000)$. For the positive branch (left panel) the maximum amplitude of $\delta$ today is $\pm 2\cdot 10^{-4}$. Thus, the final amplitude is 20 times smaller than the initial ones guaranteeing that $\delta$ remains in the linear regime and the perturbations remain small. The final value for $\delta$ turns out to be independent of the initial conditions for $\delta$ and does depend solely upon the initial conditions of $\delta_m$. Concerning $\delta_m$, its final amplitude for $\delta_m$ acquires maximum-minimum values of $\pm2\cdot10^{-3}$, i.e., twice times the initial amplitude. These values depend both upon initial conditions for $\delta$ and $\delta_m$. For the negative branch (right panel) the maximum amplitude of $\delta$ today is $\pm2\cdot10^{-3}$. The value for $\delta_m$ today depends both upon the initial conditions fixed for $\delta$ and $\delta_m$. The maximum amplitude for $\delta_m$ is $\pm0.04$, i.e, 40 times bigger than the initial amplitude.}}
	
%
%
%
	\label{fig:LCDM_Valores_plus-minus_today}
\end{figure*}

\begin{figure*}[h!]
	\centering
		\includegraphics[width=0.4775\textwidth]{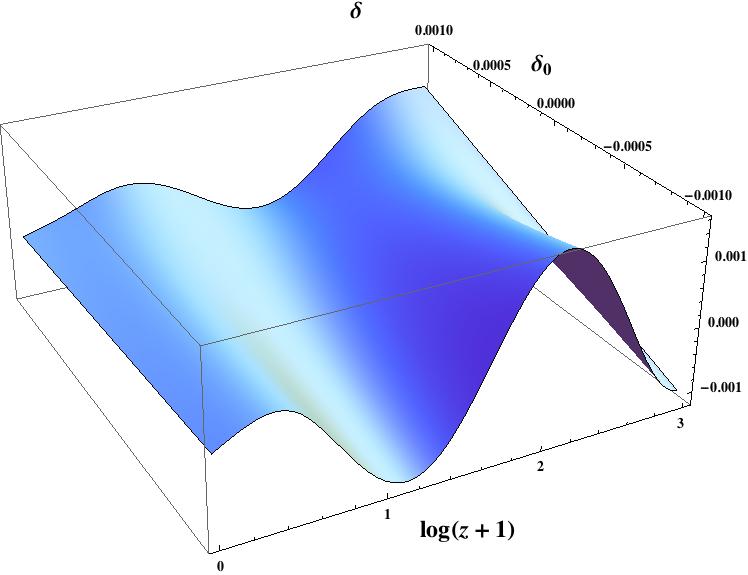}\,\,\,\,\,\,\,\,
		\includegraphics[width=0.4775\textwidth]{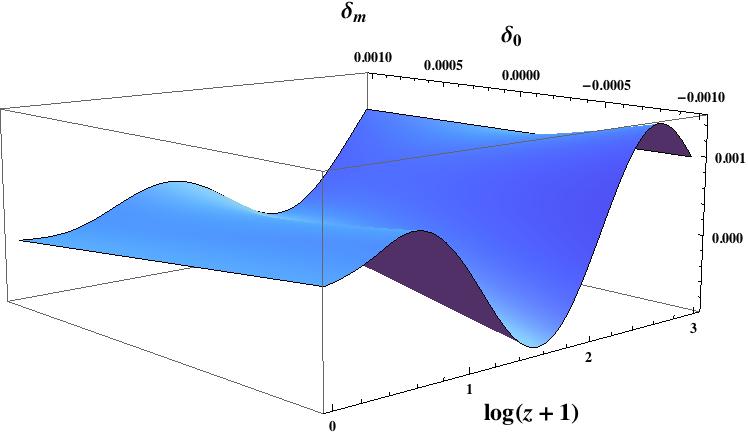}
		 \includegraphics[width=0.4775\textwidth]{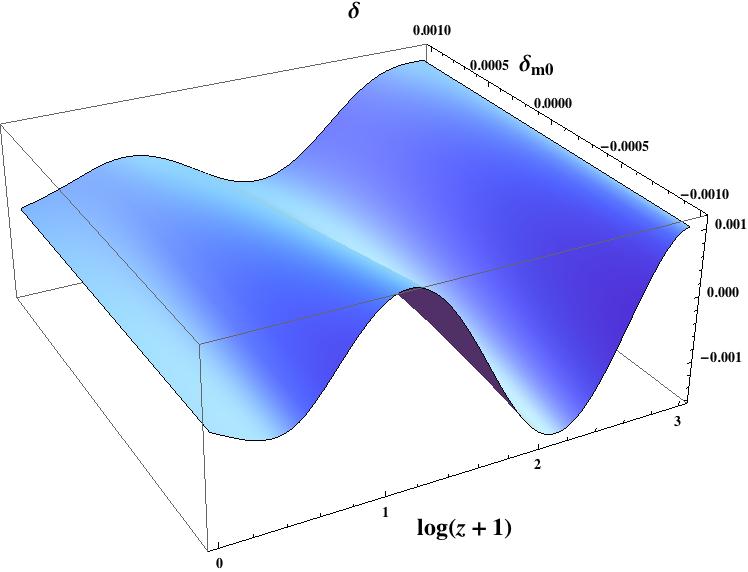}\,\,\,\,\,\,\,\,
		\includegraphics[width=0.4775\textwidth]{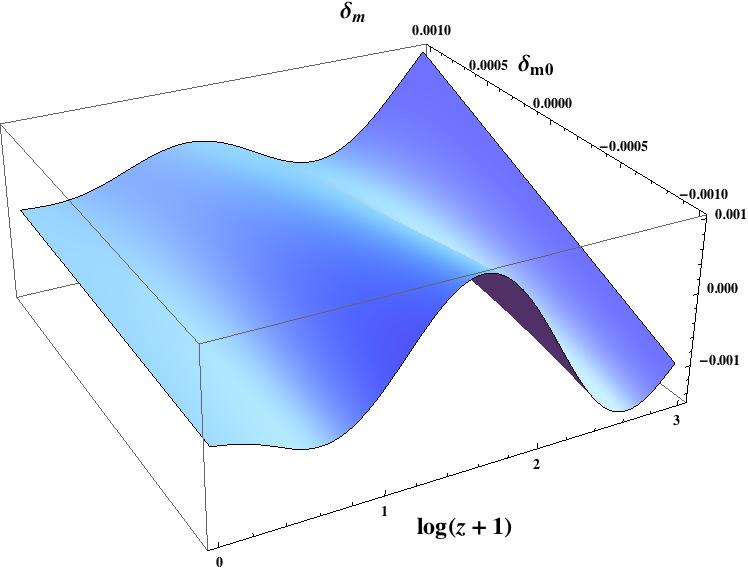}	
		%
%
		\caption{\footnotesize{
Stability of $f(R,G)$ given by (\ref{model_LCDM}) for positive sign in expression (\ref{GBLCDM}):
Evolution in redshift of cosmological and dust perturbations, $\delta$ and $\delta_m$ respectively. 
Different initial conditions at redshift $z=1000$ were imposed. 
The two upper panels represent the evolution in redshift for $\delta$ (left) and $\delta_m$ (right) for 
fixed $\delta_m(z=1000)=10^{-3}$  whereas $\delta(z=1000)$ varied from $-0.001$ to $0.001$. 
On the left panel, one can see that regardless the initial conditions  for $\delta$ this quantity today approaches 
null amplitude with decreasing oscillating amplitude. On the right panel, $\delta_m$ follows the same behavior decaying from $10^{-3}$ 
to null amplitude today.
At the lower figures, $\delta$ (left) and $\delta_m$ (right)  are depicted again this time for 
fixed $\delta(z=1000)=10^{-3}$  whereas $\delta_{m}(z=1000)$ varied from $-0.001$ to $0.001$. 
The oscillatory character of the upper figures appears again. In fact, in both panels of the lower figures, $\delta$ tends to zero 
regardless the initial value of $\delta_m$ as well as does $\delta_m$ today.
		}}
	\label{fig:LCDM_ValoresRedshift_plus}
\end{figure*}

\begin{figure*}[h!]
	\centering
		\includegraphics[width=0.4775\textwidth]{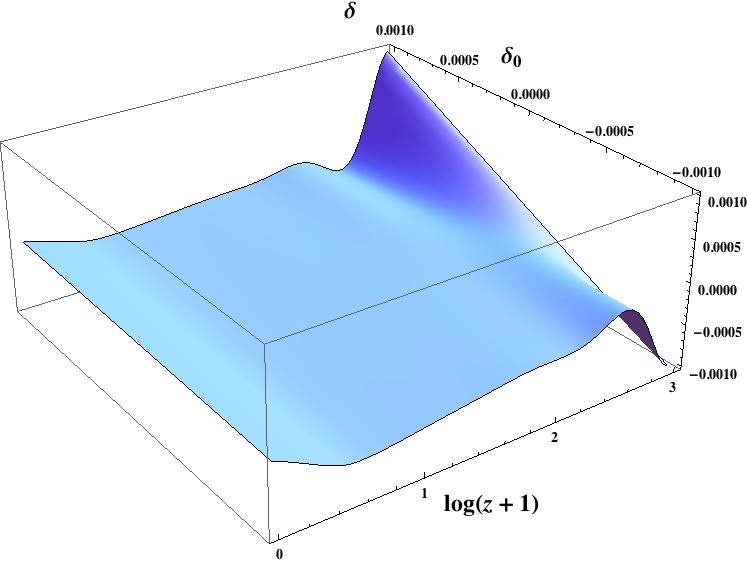}\,\,\,\,\,\,\,\,
		\includegraphics[width=0.4775\textwidth]{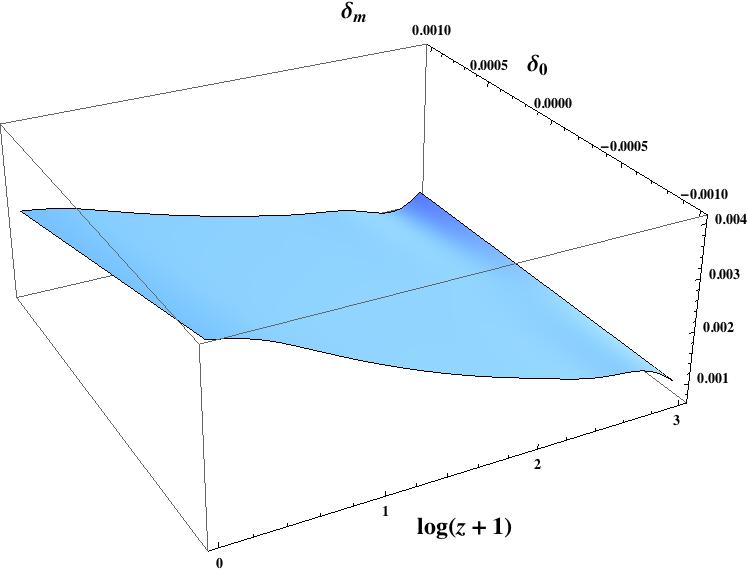}
		 \includegraphics[width=0.4775\textwidth]{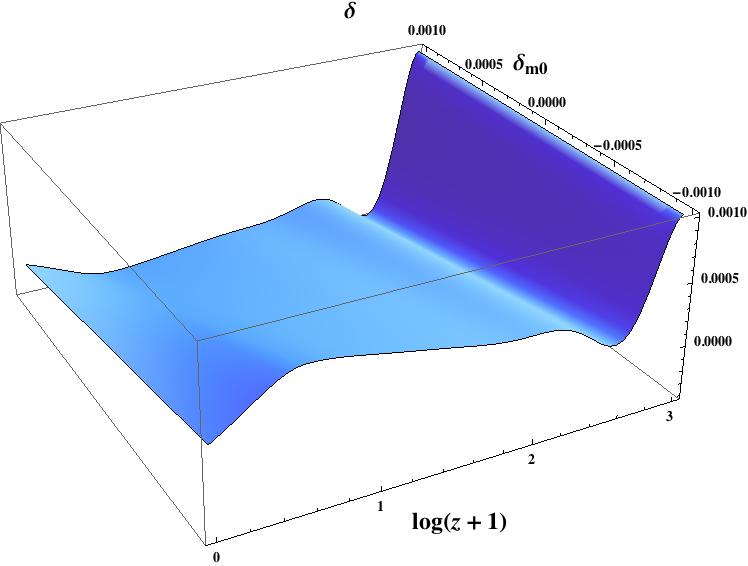}\,\,\,\,\,\,\,\,
		\includegraphics[width=0.4775\textwidth]{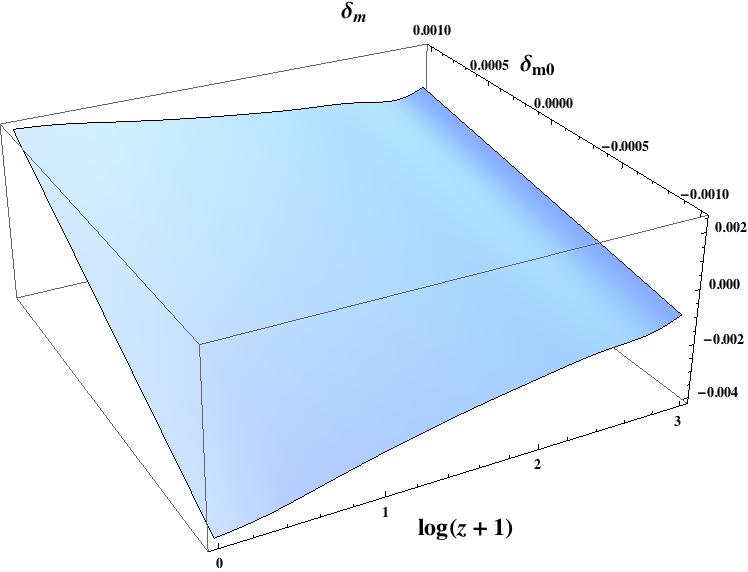}		
		%
		%
%
		\caption{\footnotesize{
Stability of $f(R,G)$ given by (\ref{model_LCDM}) for negative sign in expression (\ref{GBLCDM}):
Evolution in redshift of cosmological and dust perturbations, $\delta$ and $\delta_m$ respectively. Different initial conditions at redshift $z=1000$ were imposed.  The two upper panels represent the evolution in redshift for $\delta$ (left) and $\delta_m$ (right) for  fixed $\delta_m(z=1000)=10^{-3}$  whereas $\delta(z=1000)$ varied from $-0.001$ to $0.001$. 
On the left panel, one can see that regardless the initial conditions  for $\delta$ this quantity today approaches 
null amplitude. On the right panel, $\delta_m$ instead grows to bigger amplitudes achieving today values between $2-4\cdot10^{-3}$. At the lower figures, $\delta$ (left) and $\delta_m$ (right)  are depicted again this time for  fixed $\delta(z=1000)=10^{-3}$  whereas $\delta_{m}(z=1000)$ varied from $-0.001$ to $0.001$. 
$\delta$ tends to zero amplitude regardless the initial value of $\delta_m$. Nonetheless, $\delta_m$ increases its final amplitude today ranging from  $-4\cdot10^{-3}$ to $2\cdot10^{-3}$ with regard to the initial amplitude.}}
	\label{fig:LCDM_ValoresRedshift_minus}
\end{figure*}

Now that we have revised the form of the $f(R,G)$ Lagrangian able to mimic the background solution (\ref{a_LCDM}), we study the 
system made of by the equations (\ref{2.5}) and (\ref{2.6}). 
The stability of this model for several different initial conditions can be established
has been represented in Figs.  \ref{fig:LCDM_Valores_plus-minus_today},  \ref{fig:LCDM_ValoresRedshift_plus} and \ref{fig:LCDM_ValoresRedshift_minus}.
Both signs in expression (\ref{GBLCDM}) were considered leading to two different analyses: 
For the positive branch in expression (\ref{GBLCDM}), the obtained solutions were  oscillatory with decreasing amplitude for all the studied initial 
conditions. According to the numerical results depicted in Figures  \ref{fig:LCDM_Valores_plus-minus_today} (left panel) and  \ref{fig:LCDM_ValoresRedshift_plus},  
$\delta_{m}(t)$ attains bigger amplitudes today (in absolute value) than $\delta(t)$ in this branch. 
%
%
%
On the other hand, for the negative branch in expression (\ref{GBLCDM}), the $\delta$ decrease in amplitude whereas $\delta_m$ solutions increase 
for all the studied initial conditions.
According to Figure  \ref{fig:LCDM_Valores_plus-minus_today})  (right panel), 
$\delta_m$ increases in absolute value faster than $\delta$. Fig. \ref{fig:LCDM_ValoresRedshift_minus} 
illustrates how $\delta$ amplitude tends to decay whereas $\delta_m$ increases.

\newpage
\newpage


\section{Conclusions}

In this work we have extended the study of $f(R,G)$ modified gravity theories to provide an accurate description of the instabilities that these theories may present.
For such scenarios, the linearized perturbed equations have been derived. They were obtained once the modified Einstein equations are implemented with perturbations in both the Hubble parameter and matter density. The resulting coefficients have been explicitly presented  for the first time in the existing literature and a strong dependence on the chosen $f(R,G)$ model was observed.

We have studied three of the most important cosmological solutions in the standard cosmological concordance model around a spatially flat FLRW background: de Sitter expansion, power laws and the scale factor solution as provided for $\Lambda$CDM model. The required $f(R,G)$ gravitational Lagrangians to provide de Sitter and power-law solutions are explicitly determined whereas the model mimicking $\Lambda$CDM is revised . Concerning the perturbations, in the first case, we have found that a certain combination in the sign of some derivatives of the gravitational Lagrangian evaluated in the cosmological background is required to ensure the stability of the solution. This may provide a way to understand  the end of an inflationary era produced by one of these de Sitter points (which appear as natural solutions of $f(R,G)$ gravity), as well as for understanding the evolution of dark energy epoch. \\

With respect to the power-law solutions, the complexity of the coefficients for the perturbed equations
led us to study two representative models:  either a sum of functions of scalar curvature and Gauss-Bonnet term or powered 
products of these two scalars. These models  may encapsulate the main features for 
cosmologically viable $f(R,G)$ theories in some asymptotic regimes. 
Their rich phenomenology was summarized in Table \ref{Table_II}. 

Besides, the usual power-law behavior for the scale factor in general relativity when either dust or radiation are the dominant components in the fluid sector, was found to be mimicked, even in the absence of such fluids, by appropriate choices in the parameter space of these models.  
The radiation dominated evolution, i.e. $a(t)\propto t^{1/2}$, deserves special attention in the case of actions of the type $f(R,G)=f(G)+f(R)$. For this case, we have shown that the background evolution requires the action reduces to Hilbert-Einstein action, a result that was also naturally extended to $f(R)$ theories. Consequently, the radiation dominated era can not be described in principle for this kind of Lagrangians, or at least,  effects of such terms should be negligible during that epoch.  
Concerning the evolution of the perturbations, we have shown that for models of the form $f(R,G)=f_1(G)+f_2(R)$, the perturbations oscillate tending to zero at $z=0$ provided that the initial condition $\delta_{0}'$ is assumed to be null, and the background evolution ($\alpha$) coincides with the EoS parameter of the perfect fluid. 
Also note that independently of the model, the final values has a direct linear relation with the initial ones imposed. Note also that in the case $\alpha=2$ (accelerating expansion), the perturbed equation presents a pole at a particular redshift, while the perturbations remain small up to the divergence.  \\

Finally, the case of the $f(R,G)$ model able to mimic the $\Lambda$CDM scale factor led us as well to relevant conclusions. According to previous literature, this model presents two branches. Our analysis showed the decreasing and oscillatory character of perturbations in the positive branch regardless the initial conditions. Therefore, this solution may be considered as stable with respect to small perturbations. On the other hand, the negative branch showed that matter perturbations are not  constrained in amplitude even if the perturbations for the Hubble parameter approach zero for a wide choice of initial conditions. 

Hence, the wide study on both cosmological solutions and stability of homogeneous perturbations for in $f(R,G)$ gravity carried out in this paper may help to a better understanding for such higher order theories of gravity. We have shown that gravitational action plays a very important role in the stability of the solutions depending both upon the form of the  $f(R,G)$ theory and the parameters of the model, which means a great  difference with respect to general relativity. The search for stability in widely accepted cosmological solutions helps to constrain the parameters space of the $f(R,G)$ models that may be viable and therefore may deserve further study 
by future analyses of the full spectrum of cosmological perturbations in modified gravity.

\begin{table}[h]
\begin{center}
\begin{tabular}{|c|c|c|}
\hline
\hline
\multicolumn{3}{|c|}{Model $f(R,G)=f_1(G)+f_2(R)$} \\
\hline
\hline
  $a(t)\propto t^{\alpha}$ & EoS & Stability \\
\hline
\multirow{2}{*}{$\frac{2}{3}$}  & $w_m=0$ &  Stable with $\dot{\delta}_0=0$. $\delta_m$ grows very fast when  $\dot{\delta}_0\neq0$. \\
 & $w_m=1/3$ & Unstable: $\delta$ and $\delta_m$ grow with $z$.  \\
\hline
$2$  & $w_m=0$  & Existence of a pole $z_{pole}$ fixed at large redshift. Stable close to $z=0$. \\
\hline
\end{tabular}
\end{center}
\begin{center}
\begin{tabular}{|c|c|c|c|}
\hline
\hline
\multicolumn{4}{|c|}{Model $f(R,G)=\mu R^{\beta}G^{\gamma}$} \\
\hline
\hline
Configuration& $a(t)\propto t^{\alpha}$ & EoS & Stability \\
\hline
\multirow{2}{*}{Vacuum} & $\alpha=\frac{1}{2}$ & - & $\beta>1$,   $\gamma$ one or even number and $\gamma<\frac{1}{8}\left(5-4\beta\right)$ \\
& $\alpha=\frac{2}{3}$   & -  & Any $\beta$ $\gamma$ one or even number and $\gamma=\gamma_{-}$ given by eqn. (\ref{gammas}) \\
\hline
\multirow{5}{*}{Fluid} & $\alpha=\frac{1}{2}$  & $\omega=\frac{1}{3}$ & Impossible to satisfy. \\
 & $\alpha=\frac{2}{3}$  & $\omega=0$  & Eqns. (47) and (48) must be satisfied,  \\
& &  & $\delta_m$ decreases and $\delta$ increases. \\
& $\alpha\neq\frac{1}{2}$  & $\omega= \frac{1}{3}$  & Impossible to satisfy. \\
 & $\alpha=\frac{4}{3}$  & $\omega= 0$  & $\beta=-1$, $\gamma=\frac{3}{2}$, $\delta$ tends to zero and $\delta_m$ increases. \\
\hline
\end{tabular}
\end{center}

\caption{
\footnotesize{Summarization of the stability of power-law solutions for $f(R,G)$ models given by expressions (\ref{2.9}) and (\ref{product_power_laws}). In the cases where no fluids are considered (Vacuum) it can be seen that stability of the cosmological solutions can be achieved for appropriate choice in the parameters space. Once either dust of radiation fluids are considered, one sees that usual power-law evolution is on the one hand not feasible for radiation and, on the other hand, the stability depends on the initial conditions for dust. By the a particular action for this kind of cosmological solutions, modified gravity may contribute during the matter/radiation dominated eras, but also reproduce dark energy epoch($\alpha=2$)}.} 
\label{Table_II}
\end{table}

%
\ack
We would like to thank Sergei D. Odintsov and Antonio L\'opez Maroto for useful discussions on the topic. AdlCD acknowledges financial support from URC (South Africa) and MICINN (Spain) projects numbers FIS2011-23000, FPA2011-27853-C02-01 and Consolider-Ingenio MULTIDARK CSD2009-00064. DSG acknowledges support from a postdoctoral contract from the University of the Basque Country (UPV/EHU) under the program ``Specialization of research staff''. DSG is also supported by the research project FIS2010-15640, and also by the Basque Government through the special research action KATEA and UPV/EHU under program UFI 11/55.
%

%

\section*{Appendix}
In this Appendix we present explicitly the coefficients of equation  (\ref{2.5}).

\begin{eqnarray}
c_2\,&=&\,-18 H_0^2(16f^0_{GG}H_0^4+8f^0_{GR}H_0^2+f^0_{RR})\,, \nonumber\\
\nonumber\\
c_1\,&=&\,-18H_0\left\{1536H_0^8\dot{H}_0f^0_{3G}-f^0_{RR}\dot{H}_0-48H_0^6\left[f^0_{GG}+8\dot{H}_0(3f^0_{RGG}+2\dot{H}_0f^0_{3G})\right]  \right. \nonumber\\
&&-8H_0^4\left[3f^0_{RG}+2\dot{H}_0(5f^0_{GG}+18f^0_{RRG}+24\dot{H}_0f^0_{RGG})\right] \nonumber\\
&&-3H_0^2\left[f^0_{RR}+8\dot{H}_0(f^0_{RG}+f^0_{3R}+2f^0_{RRG}\dot{H}_0)\right]-384H_0^7\ddot{H}_0f^0_{3G}\nonumber\\
&&\left. -288H_0^5\ddot{H}_0f^0_{RGG}-72H_0^3\ddot{H}_0f^0_{RRG}-6H_0\ddot{H}_0f^0_{3R}\right\} \,, \nonumber\\
 \nonumber\\
\frac{1}{6}c_0\,&=&\,-18432H_0^{10}\dot{H}_0f^0_{3G}+3\dot{H}^2_0f^0_{RR}+192H^8_0\left[f^0_{GG}-24\dot{H}_0(3f^0_{RGG}+5f^0_{3G}\dot{H}_0)\right]+ \nonumber\\
&&48H_0^6\left[2f^0_{RG}-3\dot{H}_0\left(7f^0_{GG}+24f^0_{RRG}+88f^0_{RGG}\dot{H}_0+48f^0_{3G}\dot{H}^2_0\right)\right] \nonumber\\
&&+12H_0^4\left\{f^0_{RR}-4\dot{H}_0\left[7f^0_{RG}+6f^0_{3R}+3\dot{H}_0(3f^0_{GG}+14f^0_{RRG}+16f^0_{RGG}\dot{H}_0)\right]\right\} \nonumber\\
&&-H_0^2\left[f^0_{R}+3\dot{H}_0(7f^0_{RR}+8\dot{H}_0(2f^0_{RG}+3f^0_{3R}+6f^0_{RRG}\dot{H}_0))\right]-4608H^9_0\ddot{H}_0f^0_{3G} \nonumber\\
&&-3456H_0^7\ddot{H}_0(f^0_{RGG}+f^0_{3G}\dot{H}_0)-288H_0^5\ddot{H}_0(f^0_{GG}+3f^0_{RRG}+7f^0_{RGG}\dot{H}_0) \nonumber\\
&&-6H_0\ddot{H}_0(f^0_{RR}+3f^0_{3R}\dot{H}_0)-24H_0^3\ddot{H}_0\left[4f^0_{RG}+3(f^0_{3R}+5f^0_{RRG}\dot{H}_0)\right]\,, \nonumber\\
\nonumber\\
c_m\,&=&\,\kappa^2\rho_{m0}(t)\ .
\end{eqnarray}


\section*{References}


\begin{thebibliography}{99}


\bibitem{review}
S.~Nojiri and S.~D.~Odintsov,
  eConf {\bf C0602061}, 06 (2006)
  [Int.\ J.\ Geom.\ Meth.\ Mod.\ Phys.\  {\bf 4}, 115 (2007)];
  Phys.\ Rept.\  {\bf 505}, 59 (2011)
  [arXiv:1011.0544 [gr-qc]].
S.~Capozziello and M.~Francaviglia,
  Gen.\ Rel.\ Grav.\  {\bf 40}, 357 (2008);
T.~P.~Sotiriou and V.~Faraoni, Rev. Mod. Phys. \textbf{82} 451 (2010);
F.~S.~N.~Lobo,
  arXiv:0807.1640 [gr-qc]; 
S.  Capozziello, M.  De Laurentis, Physics Reports \textbf{ 509} 167-321  (2011); 
  A.~de la Cruz-Dombriz and D.~Saez-Gomez,
  Entropy {\bf 14}, 1717 (2012).

\bibitem{book} S. Capozziello, V. Faraoni, ``Beyond Einstein gravity: A Survey of gravitational theories for cosmology and astrophysics'' in Fundamental Theories of Physics, Vol. 170, Springer, 2010. (ISBN-10: 9400701640, ISBN-13: 978-9400701649)



\bibitem{Ref1} A. Dobado and  A. L. Maroto {\it Phys. Rev.} {\bf D 52} 1895 (1995).

\bibitem{Ref2} G. Dvali, G. Gabadadze and M. Porrati {\it Phys. Lett.} {\bf B 485} 208 (2000).

\bibitem{Ref3} J. Beltr\'an and A. L. Maroto {\it Phys. Rev.}  {\bf D 78} 063005 (2008); {\it JCAP} {\bf 0903} 016 (2009)
; {\it Phys. Rev.} {\bf D 80 } 063512 (2009);  {\it Int. J. Mod. Phys.} {\bf D 18}  2243-2248 (2009).






\bibitem{unification}
  S.~Nojiri and S.~D.~Odintsov,
  Phys.\ Rev.\ D\ {\bf 74} (2006) 086005
  [hep-th/0608008];

E.~Elizalde and D.~S\'aez-G\'omez,
  Phys.\ Rev.\ D\ {\bf 80}, 044030  (2009)
  [arXiv:0903.2732 [hep-th]].

S.~Nojiri, S.~D.~Odintsov and D.~S\'aez-G\'omez,
  Phys.\ Lett.\ B\ {\bf 681}, 74  (2009)
  [arXiv:0908.1269 [hep-th]].

  \bibitem{Perturbations}  
J. M. Bardeen, Phys. Rev. D22 1882 (1980);

S. M. Carroll, I. Sawicki, A. Silvestri and M. Trodden, New J. Phys. 8 323, (2006); 

Y. S. Song, W. Hu and I. Sawicki, Phys. Rev. D75 044004, (2007);

A. A. Starobinsky, JETP Lett. 86, 157 (2007);

R. Bean, D. Bernat, L. Pogosian, A. Silvestri and M. Trodden. Phys. Rev. D75 064020, (2007).

S. Carloni, P. K. S. Dunsby and A. Troisi, Phys. Rev. D77 024024, (2008); 

S. Tsujikawa, Phys. Rev. D77:023507, (2008);

S. Tsujikawa, K. Uddin, R. Tavakol, Phys. Rev. D77:043007, (2008); 

A. de la Cruz-Dombriz, A. Dobado and A. L. Maroto, Phys. Rev. D77 123515 (2008); Phys. Rev. Lett. D 103, 179001 (2009); 

A.~Abebe {it et al.}, 
  Class.\ Quant.\ Grav.\  {\bf 29}, 135011 (2012). 
  

\bibitem{BH}
S. Mignemi and D. L. Wiltshire, Phys. Rev. D 46, 1475 (1992);

M. Cvetic, S. Nojiri and S. D. Odintsov, Nucl. Phys. B 628, 295 (2002);

R. G. Cai, Phys. Rev. D 65, 084014 (2002);

Y. M. Cho and I. P. Neupane, Phys. Rev. D 66, 024044 (2002);
G. Cognola, E. Elizalde, S. Nojiri, S. D. Odintsov and S. Zerbini, JCAP 0502, 010 (2005);

R. G. Cai, Phys. Lett. B 582, 237 (2004); J. Matyjasek, M. Telecka and D. Tryniecki, Phys. Rev. D 73, 124016 (2006);

T. Multamaki and I. Vilja, Phys. Rev. D 74, 064022 (2006);

G. J. Olmo, Phys. Rev. D 75, 023511 (2007);

D. N. Vollick, Phys. Rev. D 76, 124001 (2007);

F. Briscese and E. Elizalde, Phys. Rev. D 77, 044009 (2008);

A. M. Nzioki, S. Carloni, R. Goswami and P. K. S. Dunsby, Phys. Rev. D81 084028 (2010);

S. Capozziello, M. De Laurentis and A. Stabile, Class. Quant. Grav. 27, 165008 (2010);

Y. S. Myung, Phys. Rev. D84 024048 (2011).

\bibitem{Ref5} A. A. Starobinsky {\it Phys. Lett.} {\bf 91B} 99 (1980);

S. Nojiri and S. Odintsov {\it Phys. Rev. } {\bf D 68} 123512 (2003);
 {\it Gen. Rel. Grav.}  {\bf 36} 1765 (2004);

S. Capozziello {\it Int. J. Mod. Phys.} {\bf D 11}, 483 (2002);

S. M. Carroll {\it et al}, {\it Phys. Rev.} {\bf  D71} 063513 (2005);

S. Carloni, P. K. S. Dunsby, S. Capozziello and A. Troisi 
{\it Class. Quant. Grav.} {\bf 22} 4839 (2005). 

T. P. Sotiriou and S. Liberati 
{\it Annals of Physics} 332 (2007);

J. A. R. Cembranos {\it Phys. Rev.} {\bf D 73} 064029  (2006);

 T. Clifton and J. D. Barrow 
{\it Phys. Rev.} {\bf D 72} 103005  (2005). 

S.~Capozziello, S.~Carloni and A.~Troisi,
  Recent Res.\ Dev.\ Astron.\ Astrophys.\  {\bf 1}, 625 (2003)
  [astro-ph/0303041];
  
    J.~A.~R.~Cembranos,
  Phys.\ Rev.\ Lett.\  {\bf 102} 141301 (2009)
  [arXiv:0809.1653 [hep-ph]];
  
  D.~S\'aez-G\'omez,
  Gen.\ Rel.\ Grav.\  {\bf 41}, 1527 (2009)
  [arXiv:0809.1311 [hep-th]];
  
    P.~K.~S.~Dunsby, E.~Elizalde, R.~Goswami, S.~Odintsov and D.S\'aez-G\'omez,
  Phys.\ Rev.\ D {\bf 82}, 023519 (2010)
  [arXiv:1005.2205 [gr-qc]].

\bibitem{Ref6}
V. Faraoni 
({\it Preprint} arXiv:0810.2602v1 [gr-qc]) (2008).   


\bibitem{gr-qc/0607118} 
  A.~de la Cruz-Dombriz and A.~Dobado,
  Phys.\ Rev.\ D\ {\bf 74}, 087501  (2006)
  [gr-qc/0607118].

 \bibitem{Viable}
 
  W.~Hu and I.~Sawicki,
  Phys.\ Rev.\ D {\bf 76}, 064004 (2007)
  [arXiv:0705.1158 [astro-ph]];
  
  S.~Nojiri and S.~D.~Odintsov,
  Phys.\ Rev.\ D {\bf 77}, 026007 (2008)
  [arXiv:0710.1738 [hep-th]];
  
     L.~Pogosian and A.~Silvestri,
  Phys.\ Rev.\ D {\bf 77}, 023503 (2008)
   [arXiv:0709.0296 [astro-ph]];
  
S.~Capozziello and S.~Tsujikawa,
  Phys.\ Rev.\ D {\bf 77}, 107501 (2008)
  [arXiv:0712.2268 [gr-qc]].

\bibitem{BH_Dombriz} 
A. de la Cruz-Dombriz, A. Dobado and A. L. Maroto, Phys. Rev. D 80, 124011 (2009) [Erratum: Phys. Rev. D 83, 029903(E) (2011)];  J.\ Phys.\ Conf.\ Ser.\  {\bf 229} (2010) 012033;
J. A. R. Cembranos, A. de la Cruz-Dombriz and  P. Jimeno Romero, e-Print: arXiv:1109.4519 [gr-qc];   AIP Conf.\ Proc.\  {\bf 1458}, 439 (2011)  [arXiv:1202.0853 [gr-qc]].
J.~A.~R.~Cembranos, A.~de la Cruz-Dombriz and B.~Montes Nunez, JCAP {\bf 1204}, 021 (2012) [arXiv:1201.1289 [gr-qc]]; AIP Conf.\ Proc.\  {\bf 1458}, 491 (2011).
 



\bibitem{Cog1}
G. Cognola, E. Elizade, S. Nojiri, S. D. Odintsov and S. Zerbini, 
Phys. Rev. D \textbf{73} 084007 (2006) [arxiv:hep-th/0601008].

\bibitem{arXiv:1001.3636}
  E.~Elizalde, R.~Myrzakulov, V.~V.~Obukhov and D.~S\'aez-G\'omez,
  Class. Quant. Grav. \textbf{27} (2010) 095007
  [arXiv:1001.3636 [gr-qc]].

\bibitem{arXiv:1009.0902}
  R.~Myrzakulov, D.~S\'aez-G\'omez and A.~Tureanu,
  Gen.\ Rel.\ Grav.\ \ {\bf 43} (2011) 1671
  [arXiv:1009.0902 [gr-qc]].

\bibitem{Felice_malo} 
 A.~De Felice and T.~Suyama,
JCAP {\bf 0906}, 034 (2009); 
Phys.\ Rev.\ D {\bf 80}, 083523 (2009);
Prog.\ Theor.\ Phys.\  {\bf 125}, 603 (2011);
A.~De Felice, J.~-M.~Gerard and T.~Suyama,
Phys.\ Rev.\ D {\bf 82}, 063526 (2010)

\bibitem{arXiv:0810.5712} 
  A.~De Felice and S.~Tsujikawa,
  Phys.\ Lett.\ B\ {\bf 675}, 1  (2009)
  [arXiv:0810.5712 [hep-th]].
  
  \bibitem{arXiv:0911.1811} 
  A.~De Felice, D.~F.~Mota and S.~Tsujikawa,
  Phys.\ Rev.\ D\ {\bf 81}, 023532  (2010)
  [arXiv:0911.1811 [gr-qc]].
  
\bibitem{Felice_peor} A. De Felice and T. Tanaka, Prog. Theor. Phys. {\bf 124}, 503 (2010). 

\bibitem{N2}

S.~Nojiri, S.~D.~Odintsov, 
Phys. Lett. B \textbf{631} 1 (2005)  [arxiv:hep-th/0508049]; Phys. Rev. D \textbf{68}, 123512 (2003) [hep-th/0307288];
 
S.~Nojiri, S.~D.~Odintsov, M.~Sasaki, 
Phys. Rev. D \textbf{71} 123509 (2005) [arXiv:hep-th/0504052];

S.~Nojiri, S.~D.~Odintsov, O.~G.~Gorbunova, 
J. Phys. A \textbf{39} 6627 (2006) [arxiv:hep-th/0510183];

S.~Nojiri, S.~D.~Odintsov, M.~Sami, 
 Phys. Rev. D \textbf{74} 046004 (2006) [arxiv:hep-th/0605039];

S-Y.~Zhou, E.~J.~Copeland, P.~M.~Saffin, 
JCAP \textbf{0907} 009 (2009) [arXiv:0903.4610];

N.~Goheer,  R.~Goswami, Peter~K.~S.~Dunsby P.,  K.~Ananda,
 Phys. Rev. D \textbf{79} 121301 (2009) [arXiv:0904.2559];
 
K.~Uddin,  J.~E.~Lidsey, R.~Tavakol, 
Gen. Rel. Grav. \textbf{41}2725 (2009) [arXiv:0903.0270];

C.~G.~Boehmer, F.~S.~N.~Lobo, 
Phys. Rev. D \textbf{79} 067504 (2009) [arXiv:0902.2982];
 
 M.~Alimohammadi, A.~Ghalee,
Phys. Rev. D \textbf{79} 063006 (2009) [arXiv:0811.1286];

A.~De Felice, S.~Tsujikawa,
Phys. Rev. D \textbf{80} 063516 (2009) [arXiv:0907.1830];

A.~De Felice, S.~Tsujikawa,
Phys. Lett. B \textbf{675} 1 (2009) [arXiv:0810.5712];

G.~Cognola,  E.~Elizalde,  S.~Nojiri, S.~D.~Odintsov, S.~Zerbini, 
 Phys. Rev. D \textbf{75} 086002 (2007) [hep-th/0611198];

M.~Gurses,  [arXiv:0707.0347];

 B.~Li,  J.~D.~Barrow,  D.~F.~Mota,
 Phys. Rev. D \textbf{76} 044027 (2007) [arXiv:0705.3795];
 
  S.~Nojiri, S.~D.~Odintsov, P.~V.~Tretyakov, 
Phys. Lett. B \textbf{651} 224 (2007) [arXiv:0704.2520];

K.~Bamba,  S.~D.~Odintsov, L.~Sebastiani,  S.~Zerbini,
Eur. Phys. J. C \textbf{67} 295-310  [arXiv:0911.4390];
 
 S.~Capozziello,  M.~De Laurentis,  S.~Nojiri, S.~D.~Odintsov,  
Phys. Rev. D \textbf{79} 124007 (2009) [arXiv:0903.2753];

K.~Bamba, S.~Nojiri, S.~D.~Odintsov, 
JCAP \textbf{0810} 045 (2008) [arXiv:0807.2575].


\bibitem{Barrow} J. D. Barrow and A. C. Ottewill {\it J. Phys.}  {\bf A 16 } 2757 (1983); 


\bibitem{Saltas:2010tt} 
  I.~D.~Saltas and M.~Kunz,
  Phys.\ Rev.\ D {\bf 83}, 064042 (2011)
  [arXiv:1012.3171 [gr-qc]].
  

  \bibitem{arXiv:1011.2090}
  D.~S\'aez-G\'omez,
  Phys.\ Rev.\ D\ {\bf 83}  064040 (2011)
  [arXiv:1011.2090 [hep-th]].

\bibitem{GBstability} H. M. Sadjadi, Europhys. Lett. 92: 50014, 2010 arXiv:1009.2941v2 [gr-qc]: 

\end{thebibliography}
\end{document}